\documentclass[apj]{emulateapj}

\usepackage{graphicx}
\usepackage{natbib}
\usepackage{color}
\usepackage{microtype}
\usepackage[backref,breaklinks,colorlinks,citecolor=blue]{hyperref} 
\usepackage[all]{hypcap} 

\newcommand{\cotwo}{\mbox{\rm CO(2-1)}} 
\newcommand{\coone}{\mbox{\rm CO(1-0)}} 
\newcommand{\hi}{\mbox{\rm H$\,$\scshape{i}}} 
\newcommand{\hii}{\mbox{\rm H$\,$\scshape{ii}}} 

\newcommand{\ha}{\mbox{\rm H$\alpha$}} 

\newcommand{\ergpers}{\mbox{erg~s$^{-1}$}}

\newcommand{\kmpers}{\mbox{km~s$^{-1}$}}
\newcommand{\Kkmpers}{\mbox{K~km~s$^{-1}$}}
\newcommand{\Kkmperspc}{\mbox{K~km~s$^{-1}$~pc$^2$}}

\newcommand{\acounits}{\mbox{\rm M$_{\odot}$ pc$^{-2}$} \mbox{(K km s$^{-1}$)$^{-1}$}}

\newcommand{\Msunperpc}{\mbox{\rm M$_{\odot}$ pc$^{-2}$}}
\newcommand{\Msun}{\mbox{$\rm M_{\odot}$}}

\newcommand{\Msunperyr}{\mbox{\rm M$_{\odot}$ yr$^{-1}$}}

\newcommand{\shortminus}{\scalebox{0.5}[1.0]{\( - \)}}

\shorttitle{Physical Properties of Molecular Clouds}
\shortauthors{Schruba et al.}
\slugcomment{Accepted for publication in \emph{The Astrophysical Journal}}

\begin{document}
\title{Physical Properties of Molecular Clouds at 2 parsec Resolution\\ in the Low-Metallicity Dwarf Galaxy NGC 6822 and the Milky Way}

\author{
\mbox{Andreas Schruba}\altaffilmark{1}, 
\mbox{Adam~K.\ Leroy}\altaffilmark{2}, 
\mbox{J.\,M.~Diederik Kruijssen}\altaffilmark{3,4},
\mbox{Frank Bigiel}\altaffilmark{5}, 
\mbox{Alberto~D.\ Bolatto}\altaffilmark{6}, 
\mbox{W.\,J.\,G.\ de~Blok}\altaffilmark{7,8,9}, 
\mbox{Linda Tacconi}\altaffilmark{1},
\mbox{Ewine~F.\ van~Dishoeck}\altaffilmark{1,10},
\mbox{Fabian Walter}\altaffilmark{4}
}

\altaffiltext{1}{Max-Planck-Institut f\"ur extraterrestrische Physik, Giessenbachstra{\ss}e 1, 85748 Garching, Germany}
\altaffiltext{2}{Department of Astronomy, The Ohio State University, 140 W 18th St, Columbus, OH 43210, USA}
\altaffiltext{3}{Astronomisches Rechen-Institut, Zentrum f\"{u}r Astronomie der Universit\"{a}t Heidelberg, M\"{o}nchhofstra{\ss}e 12-14, 69120 Heidelberg, Germany}
\altaffiltext{4}{Max-Planck-Institut f\"{u}r Astronomie, K\"{o}nigstuhl 17, 69117 Heidelberg, Germany}
\altaffiltext{5}{Institut f\"{u}r theoretische Astrophysik, Zentrum f\"{u}r Astronomie der Universit\"{a}t Heidelberg, Albert-Ueberle Str.~2, 69120 Heidelberg, Germany}
\altaffiltext{6}{Department of Astronomy, Laboratory for Millimeter-Wave Astronomy, University of Maryland, College Park, MD 20742, USA}
\altaffiltext{7}{Netherlands Institute for Radio Astronomy (ASTRON), Postbus 2, 7990 AA Dwingeloo, The Netherlands}
\altaffiltext{8}{Astrophysics, Cosmology and Gravity Centre, Univ.\ of Cape Town, Private Bag X3, Rondebosch 7701, South Africa}
\altaffiltext{9}{Kapteyn Astronomical Institute, University of Groningen, PO Box 800, 9700 AV Groningen, The Netherlands}
\altaffiltext{10}{Leiden Observatory, Leiden University, P.O.~Box 9513, 2300 RA, Leiden, The Netherlands}

\email{Email for corresponding author: schruba@mpe.mpg.de}

\begin{abstract}
We present the ALMA survey of \cotwo\ emission from the 1/5 solar metallicity, Local Group dwarf galaxy NGC~6822. We achieve high ($0.9\arcsec \approx 2$\,pc) spatial resolution while covering large area: four $250$\,pc $\times$ $250$\,pc regions that encompass ${\sim}2/3$ of NGC~6822's star formation. In these regions, we resolve ${\sim}150$ compact CO clumps that have small radii ($\sim 2{-}3$~pc), narrow line width (${\sim} 1$~\kmpers), and low filling factor across the galaxy. This is consistent with other recent studies of low metallicity galaxies, but here shown with a $15\times$ larger sample. At parsec scales, CO emission correlates with $8\,\micron$ emission better than with $24\,\micron$ emission and anti-correlates with H$\alpha$, so that PAH emission may be an effective tracer of molecular gas at low metallicity. The properties of the CO clumps resemble those of similar-size structures in Galactic clouds except of slightly lower surface brightness and CO-to-H$_2$ ratio ${\sim}1{-}2\times$ the Galactic value. The clumps exist inside larger atomic-molecular complexes with masses typical for giant molecular cloud. Using dust to trace H$_2$ for the entire complex, we find CO-to-H$_2$ to be ${\sim}20{-}25\times$ the Galactic value, but with strong dependence on spatial scale and variations between complexes that may track their evolutionary state. The H$_2$-to-\hi\ ratio is low globally and only mildly above unity within the complexes. The SFR-to-H$_2$ ratio is ${\sim}3{-}5\times$ higher in the complexes than in massive disk galaxies, but after accounting for the bias from targeting star-forming regions, we conclude that the global molecular gas depletion time may be as long as in massive disk galaxies.
\end{abstract}


\section{Introduction}
\label{introduction}

Observations show that stars form in cold, dense clouds composed of molecular (H$_2$) gas. However, our understanding of the physical processes of molecular cloud and star formation is still limited \citep[see reviews by][]{McKee07, Kennicutt12, Tan14}. In particular, our knowledge of how molecular cloud structure relates to star formation is rapidly evolving. Over the past decade, this link has been explored via detailed observations of clouds in our own Galaxy. These show that the density structure inside molecular clouds is governed by super-sonic turbulence, which creates a (column) density probability distribution function (pdf) of log-normal shape \citep[see review by][]{MacLow04}. In molecular clouds forming stars, observations also find that this pdf exhibits a power-law tail at high column densities. This tail corresponds to small, pc-sized, high (column) density gas clumps likely to collapse under their self-gravity and form a new generation of stars \citep[e.g.,][]{Kainulainen09, Rathborne14, AbreuVicente15}. We also observe that the structure of molecular clouds has an imprint on the output stellar population: the shape of the clump mass function and the stellar initial mass function are similar \citep[e.g.,][]{Alves07, Rathborne09}; there is an apparent column density threshold for high mass star formation \citep{Kauffmann10}; and the maximum cloud mass and maximum stellar cluster mass in galaxies correlate and both increase with the gas pressure \citep{Kruijssen14b}. 

The number of clouds and the diversity of physical environments and evolutionary states that can be probed in the Milky Way remain limited. With the Atacama Large Millimeter/submillimeter Array (ALMA), we can now resolve molecular cloud structure in the nearest galaxies \citep[e.g.,][]{Indebetouw13}. This allows the prospect to measure the link between galactic environment, cloud structure, and star formation beyond only the Solar Neighborhood. A main first target of such studies are low mass, low metallicity galaxies. These ``primitive'' systems are of interest because they appear so different from large spiral galaxies like the Milky Way. They have large reservoirs of atomic gas in extended distributions, with long total gas consumption times and high gas mass fractions compared to present-day large spiral galaxies. Their low abundance of metals affects their observed properties and may be expected to influence the abundance of cold gas, the structure of cold clouds, and the ability of gas to form stars. These targets are of particular interest because early galaxies share many of these properties. Although present day dwarf galaxies are not perfect analogs to early-universe systems, the first galaxies were certainly born with few metals and few stars, so that the physics that we measure in these local galaxies would have also been at play there. 

Metals, both in the gas phase and in the form of interstellar dust, should affect the structure of star-forming clouds. Gas-phase metals act as important coolants, while dust shields cloud interiors from external radiation, which would heat the gas and dissociate molecules. Interstellar dust also facilitates molecule formation via reactions on grain surfaces. Because low temperatures make the gas clouds susceptible to gravitational collapse, both cooling and shielding are important to the ability of gas to form stars. If the formation of cold, dense gas depends on the abundance of metals, then pristine (low metallicity) environments may be rendered inefficient or unable to form stars. On the other hand, recent theoretical work suggests that molecules may not be essential to form cold, dense gas because neutral atoms or ions (e.g., carbon and oxygen) can act as effective coolants in pristine gas \citep{Glover12a}. However, these regions may also host H$_2$ molecules because of the absence of dissociating radiation. In this case, the main effect of a lack of metals may be observational, rendering molecular gas hard to see or changing the balance of atomic and molecular gas in cold regions but not the overall ability of gas to form stars. An observational confirmation of this picture remains to be made.

We need observations to test how metals affect molecular cloud structure and star formation, but the lack of metals complicates the process of observing molecular clouds. In our Galaxy, molecular cloud structure is mapped by observing dust extinction, dust emission, or molecular line emission. Unfortunately, H$_2$ molecules are a very inefficient emitter in the cold interstellar medium and absorption measurements require a bright background source. Less common, but more visible molecules, most commonly carbon monoxide (CO), are used to trace H$_2$. Of course, both dust and CO are made of metals, complicating their use to trace gas in metal-poor systems. For CO the problem is even more complex, because the abundance of CO depends on shielding by dust or H$_2$ from dissociating radiation and the conditions for CO to survive differ somewhat from those for H$_2$ to survive. In the Solar Neighborhood, this is only a modest concern because dust absorbs energetic photons over a broad wavelength range. As result, CO and H$_2$ are well-mixed and CO observations provide an efficient and reliable tracer of the molecular gas. In low metallicity environments this is no longer the case. With decreasing metal and thus dust abundance, H$_2$ self-shielding becomes the primary shielding mechanism against dissociating radiation. Due to its low abundance CO cannot (effectively) self-shield and persists only in regions where H$_2$ has absorbed all dissociating radiation in the Lyman-Werner bands \citep{Wolfire10}. Thus, CO emission traces only the densest, most opaque parts of molecular clouds, while H$_2$ remains to fill most of the cloud volume. These physics are thought to give a strong metallicity dependence for the CO-to-H$_2$ conversion factor averaged over whole clouds or galaxies \citep[see review by][]{Bolatto13}.

Despite these concerns, CO remains the second most abundant molecule in metal-poor galaxies and CO emission is an indispensable tool to detect cold, dense clouds and map their structure. Other indirect tracers of H$_2$, including optical extinction, dust emission, ionized or neutral atomic lines, and other molecular lines also suffer from metallicity effects. More practically, the resolution and sensitivity of ALMA still makes CO the fastest way to map molecular cloud structure at low metallicity.

Observations do show CO emission to be faint in low metallicity galaxies. The ratio of CO emission to star formation is a strong function of metallicity, with more metal-poor galaxies showing much less CO per unit star formation than metal-rich galaxies \citep{Schruba12}. Observations of molecular clouds in the Magellanic Clouds at ${\sim} 10$~pc resolution show that their CO luminosities are much lower than those of Galactic clouds of comparable size \citep[e.g.,][]{Fukui08, Hughes10}. On the other hand, dust emission indicates significant amounts of H$_2$ gas (i.e., excess IR emission for their \hi\ mass) so that the CO-to-H$_2$ conversion factor can be orders of magnitude higher than in our Galaxy \citep[][]{Bolatto11, Leroy11, Jameson16, Shi15}. Following the physical scenario above, one popular interpretation of these observations is that CO is selectively photodissociated compared to H$_2$ over a large area in low metallicity molecular clouds; in this case, CO molecules persist only in the most opaque, densest gas clumps \citep[e.g.,][]{Pak98, Bolatto13}.

The most direct test of this scenario is to resolve the structure of individual low metallicity molecular clouds. For a long time, this was only possible in the Large and Small Magellanic Clouds (LMC and SMC with ${\sim} 1/2$ and ${\sim} 1/5$ solar abundance, respectively), and even then only at $\gtrsim 10$~pc resolution \citep[e.g.,][]{Mizuno01, Bolatto03, Wong11}. ALMA changes this, allowing $\sim$pc scale measurements of cloud structure in low metallicity galaxies throughout the Local Group.  \citet{Indebetouw13} demonstrated this capability, presenting a sub-pc-resolution view of a (small) part of the 30 Doradus region in the LMC. Recently, \citet{Rubio15} presented \coone\ observations from the Local Group galaxy WLM, which has only ${\sim} 1/8$ solar abundance. They found CO emission to originate from small ($\lesssim 4$~pc across) structures that fill only a tiny fraction of the molecular cloud area. Despite strong differences in CO morphology, \citeauthor{Rubio15} estimated that the physical properties (density, pressure, and self-gravity) of these CO-emitting structures are comparable to clumps of similar size in metal-rich clouds as observed in the Solar Neighborhood. This result argues that the star formation process and the resulting stellar population (e.g., stellar initial mass function and star cluster properties) may be only weakly affected by changing metallicity, with the main influence of metallicity to be changing the distribution of the CO tracer molecules.

\citet{Rubio15} found ten CO-emitting clumps in two molecular clouds in one galaxy and \citet{Indebetouw13} studied a single region. With the goal of a statistical measurement of the structure of CO in low metallicity clouds over a wide area, we used ALMA to map \cotwo\ emission across five star-forming complexes in the Local Group dwarf galaxy NGC 6822. These five regions contain the bulk of the ongoing star formation activity in NGC 6822. By using a set of large mosaics (260 pointings in total) at $\lambda = 1.3$~mm, we are able to cover the whole area of each complex, from cloud core to outskirts, while still achieving the highest spatial resolution (2~pc) yet reached to study cloud structure in any galaxy beyond the Magellanic Clouds. 

In this paper, we present this new ALMA survey of NGC 6822 (\autoref{sec:data}) and use it to measure the structure of the star-forming ISM at low metallicity (\autoref{sec:methodology}). Our results are presented in \autoref{sec:results}. We estimate the global CO luminosity of NGC 6822 (\autoref{subsec:global}). Then we measure the large-scale properties --- size, mass, density, and phase balance --- of the atomic-molecular complexes that host the CO emission (\autoref{subsec:complex}). Subsequently we characterize the spatial and spectral intensity distribution of CO emission from these complexes by comparing it to other PDR tracers (\autoref{subsec:pdrtracers}) and to those in Galactic molecular clouds (\autoref{subsec:pixelwise}). Finally, we derive the small-scale properties --- size, line width, mass, and gravitational boundedness --- of the CO-bright clumps in our data and compare them to comparable-size structures in WLM and our own Galaxy (\autoref{subsec:clumpwise}). We conclude by discussing these results (\autoref{sec:discussion}) and providing a brief summary (\autoref{sec:summary}).

\subsection{The Low Metallicity Dwarf Galaxy NGC 6822}
\label{subsec:ngc6822}

\autoref{t1} summarizes the global properties of NGC 6822. In many ways NGC 6822 resembles a two times less massive version of the SMC \citep[e.g.,][]{Jameson16} except that the SMC is currently undergoing a strong interaction with the LMC and Milky Way. The proximity \citep[$D = 474 \pm 13$ kpc;][]{Rich14} makes NGC 6822 an ideal target to study cloud structure and star formation at high resolution; at this distance, $1\arcsec \approx 2.3$\,pc so that ALMA easily resolves cloud sub-structure. Like other comparatively isolated dwarf irregular galaxies, NGC 6822 is rich in gas with an atomic gas mass\footnote{Throughout the paper, all gas masses include a factor of $1.36$ to account for heavy elements and literature values are re-scaled to our adopted distance where necessary.} of $M_\mathrm{atom} \approx 1.3 \times 10^8$ \Msun\ \citep{Weldrake03, deBlok06b}. This is comparable to the galaxy's stellar mass, $M_\mathrm{star} \approx 1.5 \times 10^8$ \Msun\ \citep{Madden14}, so that the gas mass fraction is ${\sim} 50\%$. NGC 6822 is actively forming stars, the star formation rate (SFR) derived from various tracers is $\mathit{SFR} \approx 0.015$ \Msunperyr\ \citep[][and references therein]{Efremova11}, giving it a specific star formation rate, $\mathit{sSFR} \approx 10^{-10}$ yr$^{-1}$, typical of a star-forming galaxy.

\capstartfalse
\begin{deluxetable}{l@{\hskip +3pt}l@{\hskip +3pt}l}
\tablecolumns{3}
\tablecaption{Global Properties of NGC\,6822\label{t1}}
\tablehead{\colhead{Property} & \colhead{Value} & \colhead{Reference}}
\startdata
Hubble Type & IB(s)m (9.8) & NED/LEDA \\
R.\,A.\,\tablenotemark{a} & 19h\,44m\,57.74s & NED \\
DEC.\,\tablenotemark{a} & $-$14d\,48m\,12.4s & NED \\
Distance & $474 \pm 13$ kpc & \citet{Rich14} \\
Systemic Vel. & $-57 \pm 2$ km\,s$^{-1}$ & \citet{Koribalski04} \\
Inclination & $60 \pm 15$ deg & \citet{Weldrake03} \\
Position Angle & $115 \pm 15$ deg & \citet{Weldrake03} \\
E(B-V)$_{\rm foregrnd}$ & $0.21$ mag & \citet{Schlafly11} \\
E(B-V)$_{\rm internal}$ & $0.0 - 0.3$ mag & \citet{Efremova11} \\
$12+{\rm log\,O/H}$ & $8.02 \pm 0.05$ dex & \citet{GarciaRojas16} \\
$R_{25}$ & $8.69$ arcmin & LEDA \\
$M_{\rm V}$ & $-15.2 \pm 0.2$ mag & \citet{Dale07} \\
$M_{\rm star}$ & $1.5 \times 10^{8}$~\Msun & \citet{Madden14} \\
$M_{\rm atom}$ & $1.3 \times 10^{8}$~\Msun & \citet{Weldrake03}  \\
$M_{\rm mol}$ & $< 1 \times 10^7$~\Msun & \citet{Gratier10a} \\
$M_{\rm dust}$\,\tablenotemark{b} & $2.9^{+2.9}_{-0.8} \times 10^5$ \Msun & \citet{RemyRuyer15} \\
GDR\,\tablenotemark{b} & $480^{+170}_{-240}$ & for above values \\
SFR(mix) & $0.015$ \Msunperyr & \citet{Efremova11}
\enddata
\tablecomments{All masses scaled according to our adopted distance.}
\tablenotetext{a}{Optical center, the {\sc Hi} dynamical center is nearby.}
\tablenotetext{b}{Dust mass derived for an amorphous carbonaceous component.}
\end{deluxetable}
\capstarttrue

Despite abundant atomic gas and signatures of high mass star formation, NGC 6822 has a modest reservoir of molecular gas. The molecular gas mass is $M_\mathrm{mol} \lesssim 1.0 \times 10^7$ \Msun\ \citep[based on IRAM \mbox{30-m} observations by][but also see \autoref{subsec:global} \& \ref{subsec:sfgas} below]{Gratier10a}. Like other low mass galaxies, NGC 6822 is poor in metals, with metallicity ${\sim} 1/5$ the Solar value\footnote{Throughout the paper, we assume a solar oxygen abundance $12 + \mathrm{log(O/H)}_\odot = 8.69$ and a total solar mass fraction of metals $Z_\odot = 0.014$ \citep{Asplund09}.} \citep[$12+\mathrm{log\,O/H} = 8.02 \pm 0.05$,][]{GarciaRojas16, HernandezMartinez09}. It is also poor in dust, with dust mass\footnote{We adopt their dust mass estimate derived from fitting the \citet{Galliano11} semi-empirical dust model and assuming an amorphous carbon composition.} $M_\mathrm{dust} \approx 3 \times 10^5$ \Msun\ \citep{RemyRuyer15}. The implied gas-to-dust ratio is $\mathit{GDR} \approx 480$, which is ${\sim} 3$ times the Solar Neighborhood value of $\mathit{GDR}_\odot = 162$ \citep{Zubko04, RemyRuyer14}, but with a factor of ${\sim} 2$ uncertainty.

\begin{figure*}[t]
\hspace{7mm} 
\includegraphics[width=0.8\textwidth]{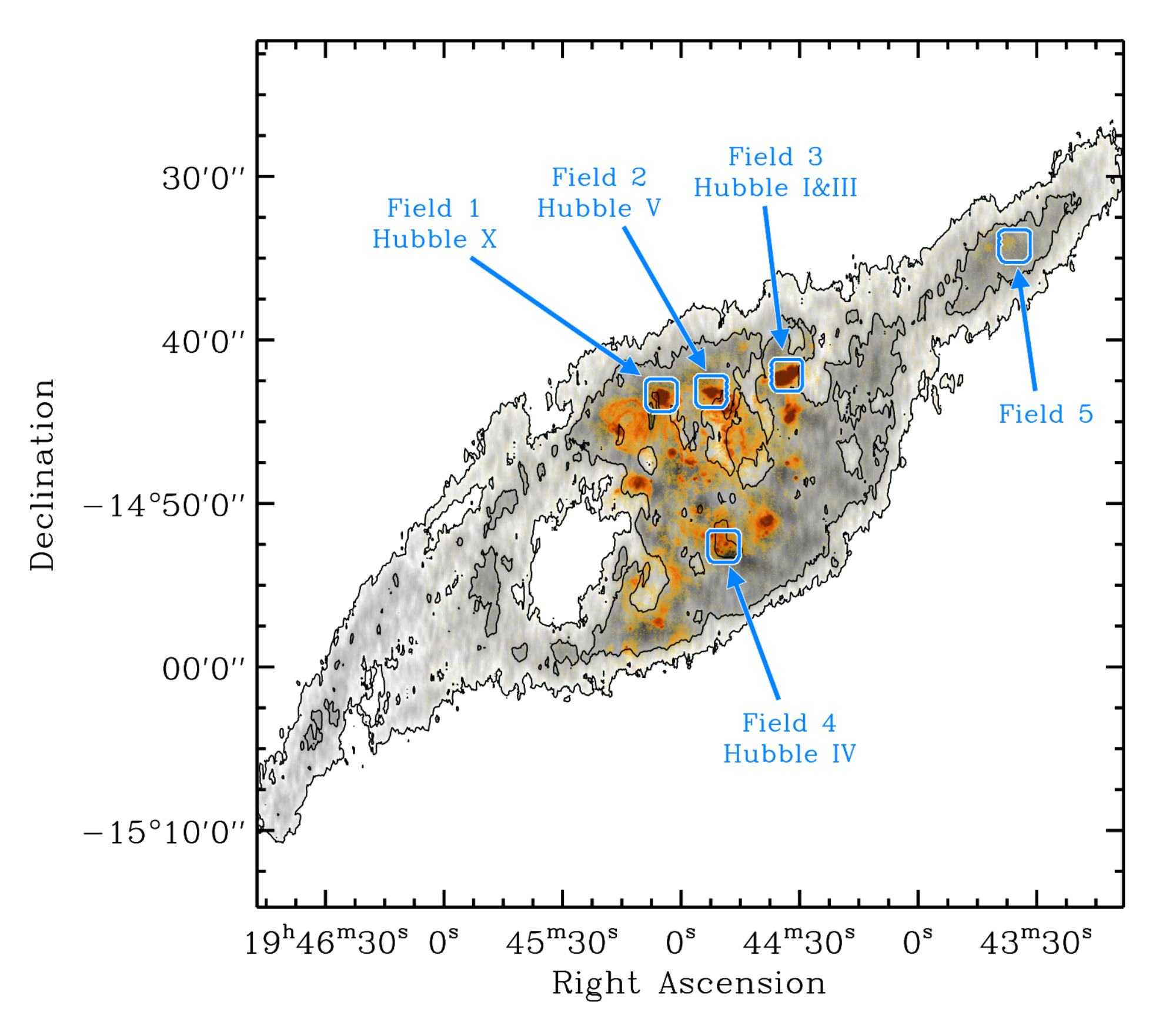}
\caption{Our five ALMA survey fields (blue rectangles, each $250\,\mathrm{pc} \times 250\,\mathrm{pc}$ in size) overlaid on an \hi\ image (grayscale) with contours at column densities of $N_\mathrm{H} = 3,~10,~30 \times 10^{20}$ cm$^{-2}$ and an \ha\ image (orange color) highlighting the location of prominent \hii\ regions. The ALMA survey covers $\sim$2/3 of NGC 6822's global H$\alpha$ and $24~\micron$ flux, implying that we map the molecular ISM hosting $\sim$2/3 of the current star formation activity. Zoom-ins for each field showing the ALMA data along with ancillary data are presented in \autoref{f2}, \ref{f3}, \ref{f4}. \label{f1}}
\end{figure*}

\autoref{f1} shows the morphology of NGC 6822 in atomic gas (gray scale) and recent star formation traced by H$\alpha$ (orange color) with our ALMA survey fields marked (blue boxes). Star formation is concentrated in the inner part of the \hi\ distribution, coincident with the main stellar disk. Our four inner ALMA fields target prominent star-forming (\hii) regions in this active area. Together, these harbor 63\% of the global H$\alpha$ flux and 65\% of the global \emph{Spitzer} $24~\micron$ flux (a tracer of embedded star formation), so that with our ALMA survey we probe the cloud complexes responsible for ${\sim} 2/3$ of the current star formation activity in NGC 6822. Our fifth field targets a region in the north-west part of the \hi\ disk, selected to search for cold gas associated with low-level star formation activity evidenced in optical, ultraviolet, and H$\alpha$ imaging \citep{deBlok03, deBlok06a}.

\capstartfalse
\begin{deluxetable*}{lllllll}
\tablecolumns{7}
\tablecaption{Properties of Target Regions\label{t2}}
\tablehead{\colhead{Property} & \colhead{Unit} & \colhead{Field 1} & \colhead{Field 2} & \colhead{Field 3} & \colhead{Field 4} & \colhead{Field 5}}
\startdata
H\,{\sc ii} Region Name & \nodata & Hubble X & Hubble V & Hubble I\,\&\,III & Hubble IV & \nodata \\
Cluster Name & \nodata & OB\,13 & OB\,8 & OB\,1\,\&\,3 & OB\,5 & \nodata \\
Cluster Mass & M$_\odot$ & $7 \times 10^3$ & $4 \times 10^3$ & \nodata & \nodata & \nodata \\
No.\ O-type stars & \nodata & 35 O$-$B2V & $>40$ & \nodata & \nodata & \nodata \\
No.\ OB-type stars & \nodata & 70 O$-$B5V & 80 O$-$B5V & \nodata & \nodata & \nodata \\
H$\alpha$ Luminosity & $10^{38}$ erg~s$^{-1}$ & 3.362 (18\%) & 4.100 (22\%) & 3.409 (18\%) & 0.941 (5.0\%) & 0.010 (0.1\%) \\
8\,$\mu$m Flux Density & micro\,Jy & 0.087 (2.3\%) & 0.211 (5.5\%) & 0.055 (1.5\%) & 0.150 (3.9\%) & \nodata \\
24\,$\mu$m Flux Density & micro\,Jy & 0.241 (9.4\%) & 0.931 (36\%) & 0.155 (6.0\%) & 0.332 (13\%) & \nodata \\
\cutinhead{ALMA Cycle\,1 \cotwo\ Data (Natural Weighting)}
Beam Major Axis & arcsec & 0.97 & 1.03 & 1.04 & 1.55 & 1.16 \\
Beam Minor Axis & arcsec & 0.71 & 0.69 & 0.69 & 0.68 & 0.74 \\
Beam Size & arcsec & 0.83 & 0.85 & 0.85 & 1.03 & 0.92 \\
\nodata & parsec & 1.90 & 1.94 & 1.96 & 2.36 & 2.13 \\
Rms Noise & milly\,Jy & 18.7 & 14.1 & 14.9 & 12.4 & 13.1 \\
\nodata & K & 0.63 & 0.45 & 0.47 & 0.27 & 0.35 \\
Sensitivity & \Kkmpers & 1.1 & 0.8 & 0.8 & 0.5 & 0.6 \\
\nodata & \Msunperpc & 4.9 & 3.5 & 3.7 & 2.1 & 2.7 \\
\nodata & \Msun & 17.5 & 13.3 & 14.0 & 11.7 & 12.3 \\
Flux & $10^3$ \Kkmpers & 15.1 & 117.6 & 24.3 & 123.5 & \nodata \\
Luminosity & $10^3$ \Kkmperspc & 1.8 & 14.0 & 2.9 & 14.7 & \nodata \\
Mass (for $\alpha_\mathrm{CO,\,MW}$) & $10^3$ \Msun & 7.8 & 60.8 & 12.6 & 63.8 & \nodata \\
\cutinhead{ALMA Cycle\,1 1.3\,mm Continuum Data (Natural Weighting)}
Beam Size & arcsec & 0.87 & 0.89 & 0.90 & 1.08 & 0.96 \\
Rms Noise & milly\,Jy & 0.19 & 0.16 & 0.17 & 0.14 & 0.13 \\
\cutinhead{Atomic-Molecular Complex Masses as derived from Herschel Dust Modelling}
\hi\ mass & $10^5$ \Msun & $3.97 \pm 0.04$ & $3.50 \pm 0.04$ & $0.76 \pm 0.04$ & $6.34 \pm 0.04$ & \nodata \\
Dust mass & $10^3$ \Msun & $3.1 \pm 2.2$ & $5.5 \pm 3.8$ & $1.7 \pm 1.1$ & $4.0 \pm 4.0$ & \nodata \\
GDR & \nodata & $420 \pm 345$ & $275 \pm 127$ & $410 \pm 215$ & $420 \pm 293$ & \nodata \\
Inferred \hi+H$_2$ mass & $10^5$ \Msun & $13.1 \pm 1.9$ & $15.0 \pm 4.2$ & $6.8 \pm 1.8$ & $16.9 \pm 1.9$ & \nodata \\
Inferred H$_2$ mass & $10^5$ \Msun & $9.1 \pm 1.3$ & $11.5 \pm 4.2$ & $6.0 \pm 1.8$ & $10.5 \pm 2.8$ & \nodata \\
Inferred $\alpha_{\rm CO}$ & \Msun\,pc$^{\shortminus 2}$\,(K\,km\,s$^{\shortminus 1}$)$^{\shortminus 1}$ & $572 \pm 93$ & $90 \pm 33$ & $235 \pm 72$ & $83 \pm 22$ & \nodata
\enddata
\tablecomments{Adopted distance $D = 474$ kpc; CO brightness temperature ratio $R_{21} = 1.0$; and CO-to-H$_2$ conversion factor $\alpha_{\rm CO} = 4.35$ \acounits. Sensitivity ($1\sigma$) determined over $5.0$ \kmpers. Percentages (given in parantheses) for H$\alpha$, $8\,\micron$, $24\,\micron$ fluxes state fractions of NGC\,6822's global fluxes.}
\end{deluxetable*}
\capstarttrue

\autoref{t2} summarizes the properties of our target regions. All four inner regions have been studied extensively and two of them, Hubble V \& X (our ALMA Fields 2 \& 1), are classic targets for extragalactic studies of young stellar clusters. As a result, they have been studied via ground-based optical broadband \citep{Bianchi01, deBlok00} and narrowband (H$\alpha$) imaging \citep{Hodge89, deBlok06a}, as well as space-based broadband \citep{Bianchi06, Bianchi12, Efremova11} and narrowband (various nebular lines) imaging \citep{ODell99}. These studies show that all four inner regions are actively forming massive stars and have likely done so for the past ${\sim} 10$ Myr. They  contain up to ${\sim} 100$ OB-type stars each and have H$\alpha$ luminosities of a few times $10^{38}$ \ergpers , which would rank them among the brightest and most massive star-forming regions in our Galaxy. There is no clear evolutionary sequence established in the literature, but Hubble IV \& V (our ALMA Fields 4 \& 2) show more compact CO and SFR morphologies than Hubble I{\footnotesize\&}III \& X (our ALMA Fields 3 \& 1) and higher ratios of embedded to exposed SFR tracer luminosities ($8~\micron$ or $24~\micron$ versus H$\alpha$; \autoref{t2}). Based on this, we argue below that these two regions (Fields 4 \& 2) may be currently more active (i.e., younger) than the others.

\section{ALMA Survey}
\label{sec:data}

\capstartfalse
\begin{deluxetable*}{cllccc}
\tablecolumns{6}
\tablecaption{ALMA Observations\label{t3}}
\tablehead{\colhead{Target Field} & \colhead{Execution Block} & \colhead{Start Time} & \colhead{No.\ of Antennas} & \colhead{Average Elevation} & \colhead{Precipitable Water Vapor} \\
\colhead{} & \colhead{} & \colhead{(UTC)} & \colhead{} & \colhead{(degrees)} & \colhead{(mm)}}
\startdata
Field\,1 & uid://A002/X7d44e7/X1d11 & 2014-03-23 10:06:34 & 32 (2 flagged) & 63 & 2.6 \\
Field\,2 & uid://A002/X7d727d/X63 & 2014-03-24 09:44:57 & 33 (1 flagged) & 60 & 1.4 \\
Field\,3 & uid://A002/X7d727d/X1d6 & 2014-03-24 10:29:44 & 33 (2 flagged) & 70 & 1.3 \\
Field\,4 & uid://A002/X7d76cc/X19aa & 2014-03-25 08:51:45 & 32 (0 flagged) & 49 & \nodata \\
Field\,5 & uid://A002/X7d76cc/X1e03 & 2014-03-25 11:22:38 & 31 (0 flagged) & 80 & 0.6
\enddata
\end{deluxetable*}
\capstarttrue

We observed five fields in NGC 6822 with ALMA in Cycle~1 using the \mbox{1.3-mm} Band~6 receivers (project code: 2013.1.00351.S; PI. A.~Schruba). Each field consists of 52 pointings distributed in a Nyquist-spaced hexagonal grid and covers a $110\arcsec \times 110\arcsec \approx 250~\mathrm{pc} \times 250~\mathrm{pc}$ area at $D=474$~kpc. The fields are centered on prominent \hii\ regions. They are shown in \autoref{f1} with their properties listed in \autoref{t2}. Our spectral setup includes one ``line'' spectral window targeting \cotwo . This window has bandwidth $0.938$~GHz with channel width $244$~kHz ($\approx 0.32$ \kmpers) and is centered at 230.612 GHz. This ``line'' spectral window covers the \cotwo\ line (rest frequency 230.538 GHz) over a velocity range of $-660$ to $+553$ \kmpers\ (LSRK), easily enough to capture all emission from NGC 6822 (systemic velocity $-48$ \kmpers\ LSRK).

\autoref{t3} reports dates and weather conditions for each observing session. Each session contains observations of a bandpass calibrator. This was the quasar J1924-2914 (flux density $\sim 3.2$ Jy at the time of observations) for Fields 1, 2, 3, 5 and J1733-1304 (flux density $\sim 1.4$ Jy at the time of observations) for Field~4. For all sessions, the phase calibrator was J1939-1525 (inferred flux density of 0.23 Jy at the time of observation). Titan was observed during three observing sessions (Fields 2, 3, 4) to set the absolute flux scale with an estimated uncertainty of $5\%$ using the Butler-JPL Horizons-2012 model (ALMA Memo \#594). The same flux scale was imposed on the other two data sets that lack observations of Titan (Fields 1, 5) by requiring the two quasars to have the same flux densities for all observations.

The data were processed in the Common Astronomy Software Applications package \citep[CASA, version 4.2.2;][]{Petry12} using the ``analyst-calibrated'' data sets created with help of the QA2 script-generator tool. The calibrated visibilities were imaged and deconvolved with the \texttt{clean} task using standard parameters. We chose an angular pixel scale of $0.15\arcsec$ and a channel width of $0.635$ \kmpers\ and imaged the data using natural weighting. We restored the deconvolved image using a single fixed elliptical Gaussian for each mosaic, so that each image has a single beam. 

The properties of the final data cubes are listed in \autoref{t2}. The average synthesized beam size has FWHM $0.9\arcsec \approx 2.0$~pc and the achieved rms brightness sensitivity is ${\sim} 0.5$~K. This translates to a $1\sigma$ surface brightness sensitivity of ${\sim} 0.9$ \Kkmpers\ over 5 \kmpers\ (about 2$\times$ the FWHM for a typical CO structure; see below). For an appropriate choice of conversion factor (see below), the implied $1\sigma$ sensitivity in molecular gas mass surface density is ${\sim} 8$ \Msunperpc\ and the $1\sigma$ point source sensitivity is ${\sim} 30$ \Msun .

The ALMA observations include baselines of $15{-}438$~m length. Thus, emission extending over scales larger than the maximum recoverable scale of about $0.6 \lambda / L_\mathrm{min} \approx 11\arcsec \approx 25$~pc is missing in our data sets. We can estimate the amount of missing flux from existing IRAM \mbox{30-m} \cotwo\ mapping at $15^{\prime\prime} \approx 36$~pc resolution \citep{Gratier10a}, which covers the ALMA Fields 1, 2, 3. However, only Field~2 has been robustly detected in the IRAM \mbox{30-m} data at a noise level of ${\sim} 50$~mK over $0.4$ \kmpers . For this field, the ALMA observations recover $(73 \pm 20)\%$ of the single-dish flux. The large uncertainty reflects the difficulty to determine the total flux in the IRAM \mbox{30-m} cube due to baseline instabilities. The spatial distribution of CO in our other fields is comparable to that in Field~2 or more compact, so we expect a similar level of flux recovery throughout the survey.

We identify genuine emission in the cubes by searching for emission peaks above $5\sigma$ in two adjacent channels which we then grow to include all neighboring pixels above $1.5\sigma$. This method is somewhat conservative and holds the potential to miss real but low signal-to-noise emission. However, the comparison to the IRAM \mbox{30-m} data suggests that the amount of signal missed has to be small.

In addition to \cotwo , we also observed three ``continuum'' spectral windows, each with 2~GHz bandwidth and 15.625 MHz channel width, centered at 229.196 GHz, 215.063 GHz, and 213.188 GHz. We  imaged these using the \texttt{clean} task with the ``multi-frequency synthesis'' (mfs) mode. The synthesized beam sizes ($0.9\arcsec \approx 2.0$\,pc) and achieved rms sensitivity (${\sim} 0.17$ mJy) are also listed in \autoref{t2}. Despite a few bright pixels, widespread continuum emission is not detected and we defer discussion of this part of the data to future work.

\begin{figure*}[t]
\epsscale{1.15}
\plotone{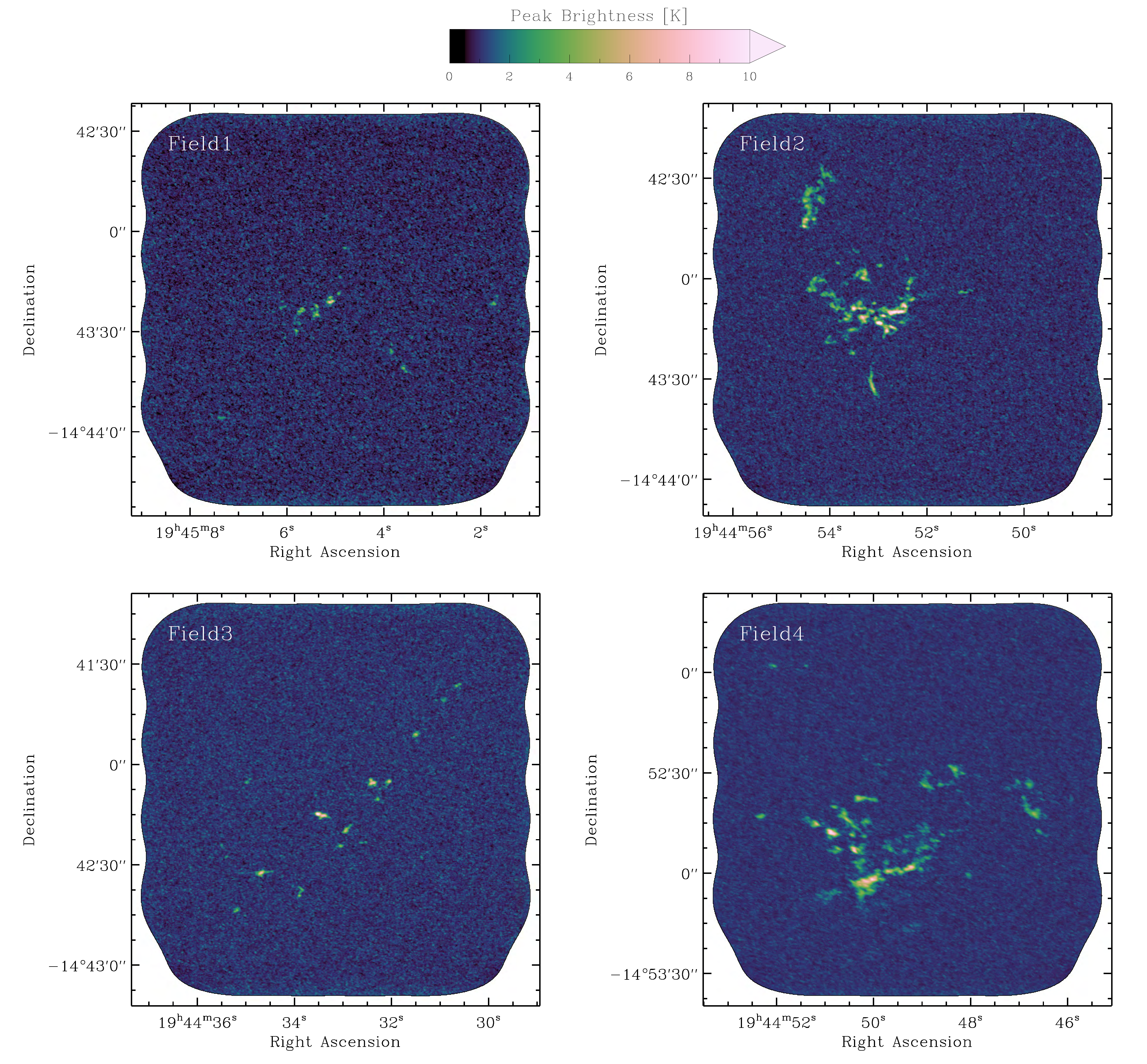}
\caption{ALMA \cotwo\ peak brightness maps for Fields $1{-}4$; Field~5 shows no genuine signal and is omitted here. The field of view of each mosaic is $110\arcsec \times 110\arcsec \approx 250\,\mathrm{pc} \times 250\,\mathrm{pc}$ and the resolution is $0.9\arcsec \approx 2.0$\,pc. The peak brightness level in signal-free regions corresponds to 2.5 times the rms noise level which varies between $0.3{-}0.6$~K among the fields (\autoref{t2}). Integrated intensity maps are shown in \autoref{f3}.
\label{f2}}
\end{figure*}

\begin{figure*}[t]
\epsscale{1.15}
\plotone{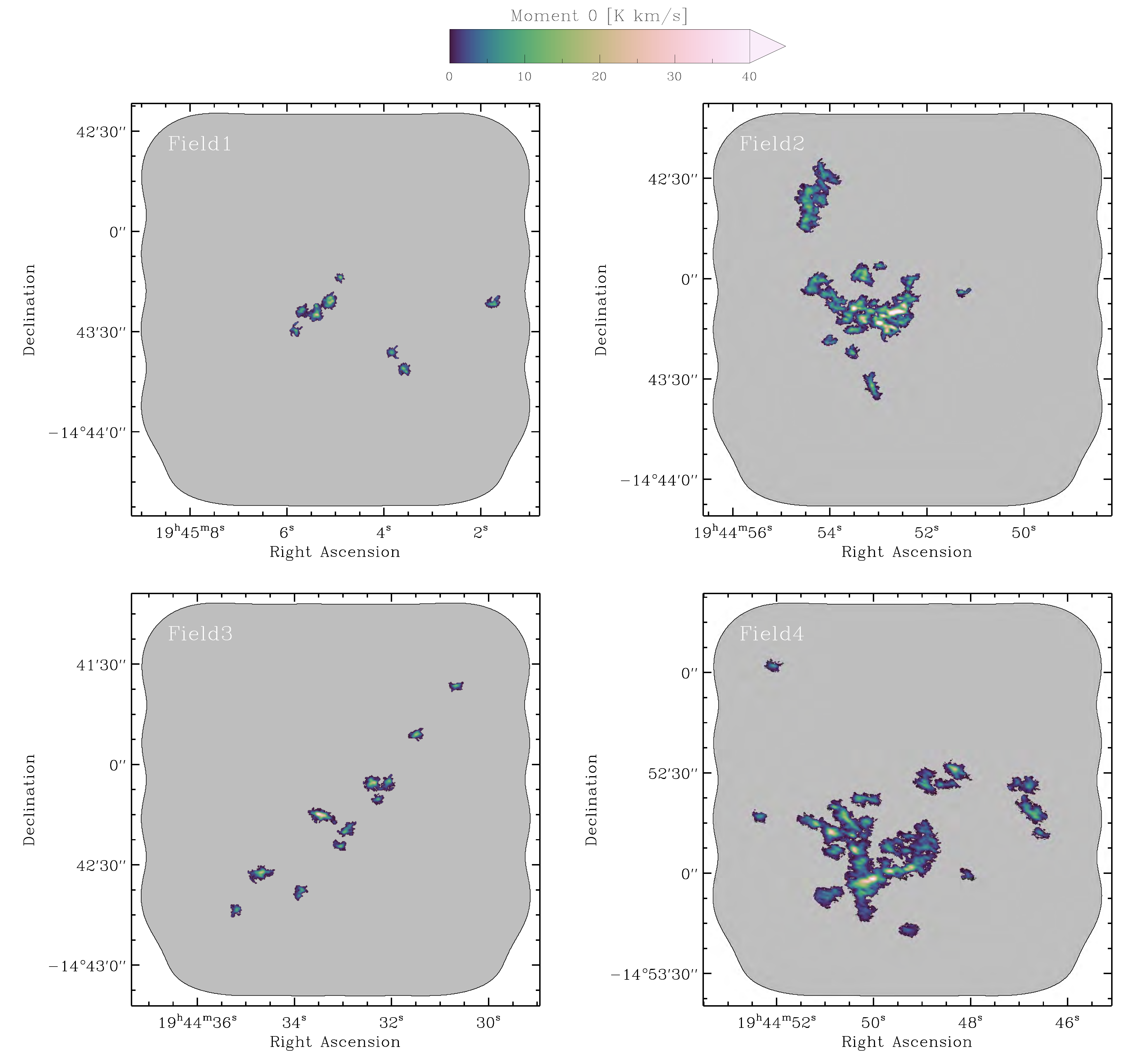}
\caption{ALMA \cotwo\ integrated intensity (moment~0) maps for Fields $1{-}4$; Field~5 shows no genuine signal and is thus omitted here. The field of view of each mosaic is $110\arcsec \times 110\arcsec \approx 250\,\mathrm{pc} \times 250\,\mathrm{pc}$ and the resolution is $0.9\arcsec \approx 2.0$\,pc. The moment maps are signal-masked (see text) and emission-free regions are shown in gray. Contour maps of the moment 0 maps with physical units are presented in \autoref{f4}.
\label{f3}}
\end{figure*}

\begin{figure*}[t]
\epsscale{1.15}
\plotone{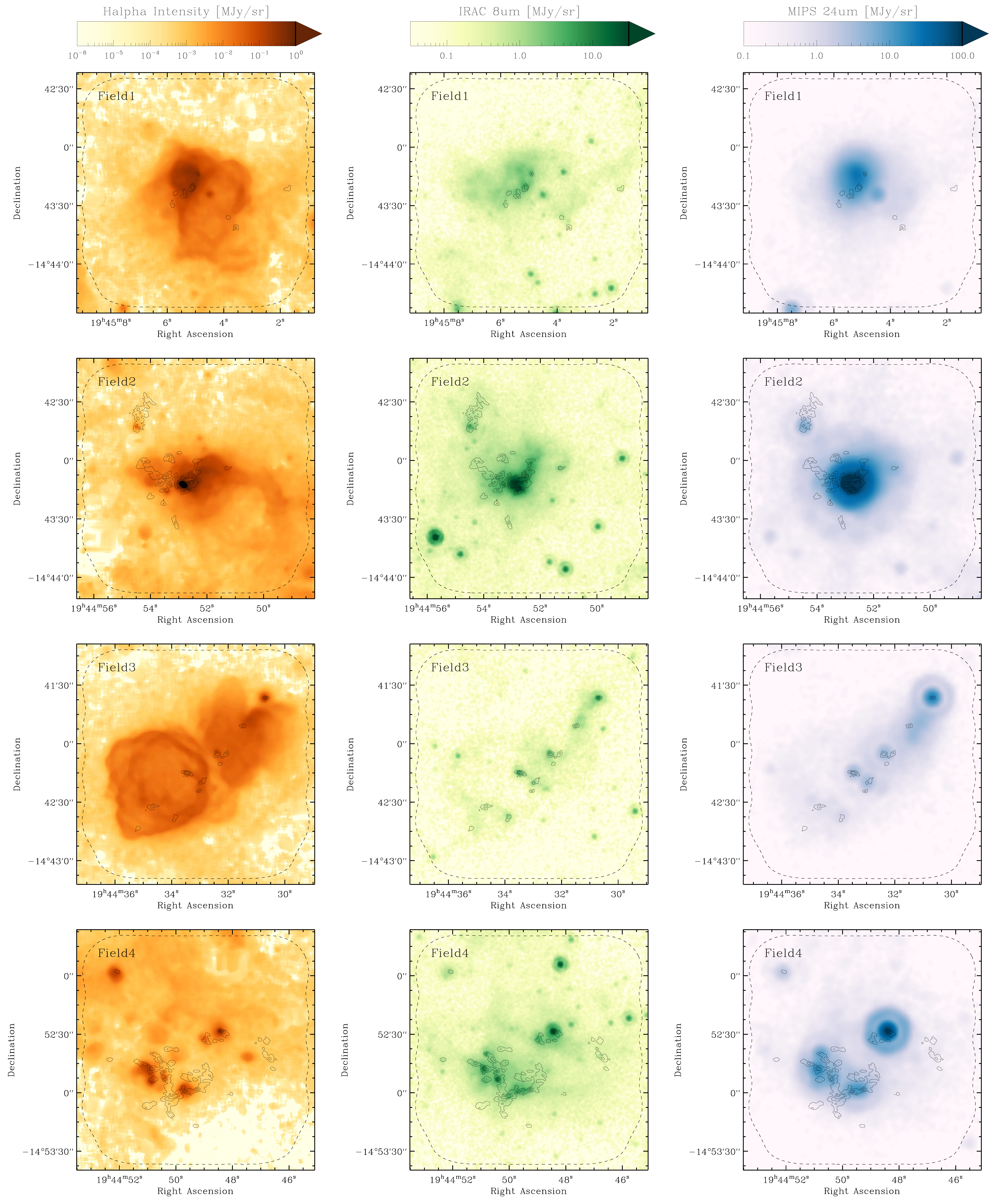}
\caption{ALMA \cotwo\ integrated intensity maps for Fields $1{-}4$ (from top to bottom) shown as contours over grayscale maps of H$\alpha$ and \emph{Spitzer} $8~\micron$ and $24~\micron$ (from left to right); the contour levels are at CO integrated intensities of $I_\mathrm{CO} = 2,~10,~20$ \Kkmpers.
\label{f4}}
\end{figure*}

\section{Other Data \& Methodology}
\label{sec:methodology}

We analyze the results of our survey in four ways. First, we extrapolate the results of our survey to make some observations about NGC 6822 as a whole. Then we consider the large-scale structure of star-forming atomic-molecular complexes, comparing the distributions of atomic gas, molecular gas, and dust across each field. Because atomic gas and dust are observed at much coarser resolution than our CO survey, this comparison is restricted to large spatial scales, reducing each of our fields to a $\sim 5 \times 5$ element grid. After this, we examine the distribution of CO at high resolution, including cross-comparison with tracers of hot dust and recent star formation. Finally, we report the detailed properties of the ${\sim} 150$ compact CO clumps in our maps.

\subsection{Tracers of Recent Star Formation}
\label{subsec:sfrtracers}

We compare the CO emission to H$\alpha$ and dust emission. At low resolution, the H$\alpha$ map \citep{deBlok06a} traces the distribution of recent star formation. At the high, $0.9\arcsec \approx 2$~pc, resolution of our ALMA data, H$\alpha$ traces the structure of {\sc Hii} regions. We compare CO to dust emission at $8~\micron$ and $24~\micron$, both from the {\em Spitzer} Infrared Nearby Galaxy Survey \citep[SINGS,][]{Kennicutt03a} and first presented in \cite{Cannon06}. The $8~\micron$ map is dominated by emission from polycyclic aromatic hydrocarbons (PAH) and traces photon dominated regions (PDR); the $24~\micron$ map measures hot dust and traces embedded star formation.

\subsection{Atomic Gas and Dust}
\label{subsec:hidust}

We use \hi\ and dust surface density maps to measure the large scale structure of the star-forming complexes. The \hi\ data are from the Very Large Array (VLA) taken as part of the LITTLE THINGS survey \citep{Hunter12}. Due to calibration complications, these were not released with the rest of the survey, however, will be presented in I.~Bagetakos et al.\ (in preparation) and have been kindly provided for use in this paper. 

The distribution of dust mass surface density is inferred from infrared data from the {\em Herschel} Dwarf Galaxy Survey \citep{Madden13}. We model the infrared spectral energy distribution (SED) between $70{-}500~\mu$m using a modified blackbody, the \citet{DraineLi07} model, and the \citet{Galliano11} model with an amorphous carbonaceous component \citep[see][for details]{Galametz10, RemyRuyer15}. The latter data set has been kindly provided by M.~Galametz and S.~Madden (private communication). The agreement between the three dust maps is generally poor. The Galactic cirrus toward NGC 6822 has similar brightness as NGC 6822 itself which severely complicates the usability of the infrared data. We have tested several methods to remove the Galactic cirrus but failed to converge on robust results. In addition, we suspect problems in the {\em Herschel} PACS and SPIRE data products themselves \citep[e.g., potentially related to the dynamic range or recovery of extended emission; see also][]{AbreuVicente16} as individual bands show inconsistencies in their intensities (by a factor of a few) with any reasonable infrared SEDs in either faint or bright regions. For that reason, we reverted using SPIRE's $350~\micron$ band as dust proxy but note that results derived from more complex dust modeling (i.e., modified blackbody, \citeauthor{Galliano11}, or \citeauthor{DraineLi07} model) agree within a factor of a few. This uncertainty dominates the error budget in our analysis of dust-inferred gas masses.

The \hi\ and dust maps are evaluated at a common resolution of $25\arcsec \approx 57$\,pc, set by the diffraction limit of SPIRE's $350~\micron$ band. At this resolution, each of our fields has $5 \times 5$ almost independent $50 \times 50$~pc elements. The outer 50 pc wide ring lacks signatures of high mass star formation and molecular gas; we use it to measure the local gas-to-dust ratio and to measure the amount of foreground and background emission from dust and \hi .

\subsection{Matched Resolution Comparison Data from the~Milky~Way~and~WLM}
\label{subsec:wlm_mw}

\capstartfalse
\begin{deluxetable}{llll}
\tablecolumns{4}
\tablecaption{Properties of Milky Way Clouds\label{t4}}
\tablehead{\colhead{Property} & \colhead{Orion} & \colhead{W3/W4} & \colhead{Carina}}
\startdata
Distance (kpc) & 0.45 & 2.0 & 2.3 \\
No.\ O-type stars & 3 & 10 & 70 \\
No.\ OB-type stars & 43 & 105 & 135 (200) \\
Cloud Mass (M$_\odot$) & $2 \times 10^5$ & $4 \times 10^5$ & $6 \times 10^5$
\enddata
\tablerefs{
\emph{Orion:} \citet{Muench08}, \citet{Wilson05};
\emph{W3/W4:} \citet{Kiminki15}, \citet{Polychroni12}; 
\emph{Carina:} \citet{Smith08}, \citet{Roccatagliata13}}
\end{deluxetable}
\capstarttrue

In order to interpret our results, we use matched resolution CO data from the Milky Way and WLM. Matched spatial resolution measurements of CO emission from Milky Way clouds offer a view on cloud structure at high (solar) metallicity, while WLM is the only other low metallicity galaxy observed so far  with data quality matching our own. The contrast of these clouds with our results in NGC 6822 illuminates how conditions in our target galaxy affect cloud structure and the degree to which conclusions about the impact of metallicity may be general (if they apply to both WLM and NGC~6822). 

In the Milky Way, we use CO maps of Orion, Carina, and W3/W4. As \autoref{t4} shows, these clouds have masses only a bit lower than the atomic-molecular complexes targeted in NGC 6822. They span a range of massive star formation activity, with Orion showing modest high mass star formation and with Carina and W3/W4 being two of the most active star-forming regions in the Galaxy. The \coone\ data for Orion and Carina are part of the CfA \mbox{1.2-m} Galactic Plane Survey\footnote{\href{https://www.cfa.harvard.edu/rtdc/CO/}{\texttt{https://www.cfa.harvard.edu/rtdc/CO/}}} and have been presented in \citet{Wilson05} and \citet{Grabelsky87}. The \cotwo\ data for the molecular cloud complex W3/W4 have been obtained with the \mbox{10-m} Heinrich Hertz Submillimeter Telescope (HHT) by \citet{Bieging11}. For a rigorous comparison, we convolve the Orion and W3/W4 data to the same spatial (2~pc) and spectral (0.635~\kmpers) resolution as our NGC 6822 data. We do not match the sensitivities, which typically are a factor of a few better for the Galactic data. The Carina data from the CfA \mbox{1.2-m} telescope have a native resolution of $5.6\,\mathrm{pc} \times 1.3\,\kmpers$; we compare these to our data at their native resolution.

We compare the clump properties that we measure for NGC 6822 to those measured from the FCRAO Outer Galaxy Survey \citep{Heyer01}. These data, as reprocessed by \citet{Brunt03}, have a resolution of $100\arcsec \times 0.98$ \kmpers . At the distance of the Perseus arm ($D \approx 2$~kpc), this corresponds to ${\sim} 1$~pc. They are thus closely matched to the resolution of our ALMA data and provide an ideal Galactic point of reference.

We also compare to the ALMA observations of WLM by \citet{Rubio15}. WLM is a Local Group dwarf galaxy with ${\sim} 1/8$ solar metallicity. Its stellar mass and current star formation activity are both ${\sim} 10$ times lower than NGC 6822. \citeauthor{Rubio15} have observed \coone\ at $6.2\,\mathrm{pc} \times 4.3\,\mathrm{pc}$ and $0.5$~\kmpers\ resolution in two atomic-molecular complexes and report the detection of ten CO-emitting structures. In our analysis of the macroscopic properties of the CO-emitting structures in NGC 6822 (\autoref{subsec:clumpwise}) we include their measurements for WLM as listed in their Table~1.

\subsection{Cloud Property Measurements}
\label{subsec:cprops}

We identify discrete objects in our data set, measure their size, line width, and luminosity, and compare them to the properties of similarly sized objects in our comparison data sets. To do this, we use an updated version of the \texttt{CPROPS} algorithm\footnote{\href{https://github.com/akleroy/cpropstoo/}{\texttt{https://github.com/akleroy/cpropstoo/}}} \citep[][and A.\,K.\,Leroy \& E.\,Rosolowsky, in preparation]{Rosolowsky06}. For details on \texttt{CPROPS}, we refer to \citet{Rosolowsky06}, \citet{Leroy15a}, and A.~Schruba et al.\ (in preparation). Briefly, we consider only significant emission, identified by a signal-to-noise cut across several channels. Within this mask, we find all significant local maxima.  For each maximum, we identify the nearby pixels that can be associated with only that peak (and no others) in an iso-intensity contour. For each region of emission, we measure its size, line width, and luminosity. We use several methods to do this: spatial and spectral moments, area measurements at half-peak value or some threshold, and ellipse fitting. Each measurement is corrected for the fact that it is made at finite sensitivity and resolution. The sensitivity calculations assume either a Gaussian profile or use a curve of growth method to account for the finite sensitivity of the data. The resolution corrections are made after the correction for sensitivity and use quadratic subtraction of the two-dimensional beam and channel width. The values reported here are the mean across all of these characterization methods. We adopt the scatter in results from different measurement approaches as our best estimate of the uncertainty of the size, line width, and luminosity, because this tends to be as large as any statistical or calibration uncertainty.

\subsection{CO Excitation and CO-to-\texorpdfstring{H$_\mathit{2}$}{H2}}
\label{subsec:excitation}

We observe \cotwo\ at high ($0.9\arcsec \approx 2$\,pc) spatial resolution. Similar to \coone, we find \cotwo\ emission to emerge from cold dense gas, especially gas with significant optical depth, though it can also be emitted under other conditions. However, the ratio of the brightness temperatures of the two lines, $R_{21}$, can vary. If \cotwo\ is sub-thermally excited or when the Rayleigh-Jeans approximation breaks down at low temperatures, then $R_{21} \leq 1$. On the other hand, gas of low opacity has $R_{21} \geq 1$. Observations in our Galaxy find $R_{21} = 0.65 \pm 0.1$ at a few 10's of parsec spatial scale; this is an average of diffuse emission from low density gas and opaque emission from dense gas \citep[][J.~Mottram et al., in preparation]{Yoda10}. Observations of nearby disk galaxies suggest a very similar line ratio of $R_{21} = 0.7$ measured on $\sim$kpc spatial scale \citep[e.g.,][]{Leroy09a,Leroy13b}. On the other hand, values of $R_{21} = 1.0 \pm 0.3$ are found in molecular clouds at 20~pc resolution in the LMC and SMC \citep{Israel03, Bolatto03} or on larger (${\sim} 100$~pc) spatial scales in IC~10 (L.~Bittle et al., in preparation). More detailed multi-transition studies of the CO emission from molecular clouds in the SMC indicate that the CO emission originates from two gas components: a more tenuous and not very dense ($n_\mathrm{H_2} = 10^2 - 10^3$ cm$^{-3}$) component of high temperature ($T_\mathrm{kin} = 100 - 300$ K) and a population of much denser clumps ($n_\mathrm{H_2} = 10^4 - 10^5$ cm$^{-3}$) of low temperature ($T_\mathrm{kin} = 10 - 60$ K) \citep{Israel03, Bolatto05}. As we will see in \autoref{subsec:clumpwise}, we do not probe the very dense clumps with our ALMA data, and most of the admittedly scarce observations of low metallicity star-forming galaxies seem to favor $R_{21} \approx 1$; therefore, we adopt $R_{21} = 1.0$ throughout the paper. This mainly affects comparisons to Galactic data and we discuss possible variations when they become relevant.

We measure CO emission but are often interested in the distribution of H$_2$. The metallicity of NGC 6822 and the high spatial resolution of our data both complicate the translation of CO to H$_2$. The CO abundance strongly depends on shielding of the dissociating radiation field and thus is a strong function of metallicity \citep{Wolfire10}. So far, the exact metallicity dependence of $\alpha_\mathrm{CO}$ remains poorly known, as does any secondary dependence on radiation field, cloud structure, and other quantities. We will derive our own estimates for $\alpha_\mathrm{CO(2-1)}$ for NGC~6822 using alternative ISM tracers (dust) and dynamical methods. Doing so, we reference to the commonly adopted Milky Way value, $\alpha_\mathrm{CO(1-0)} = 4.35$ \acounits, which includes a factor of $1.36$ to account for heavy elements \citep{Bolatto13}.

Because of our high resolution, the scale-dependence of the CO-to-H$_2$ conversion factor will also be relevant. The conventional extragalactic definition of $\alpha_\mathrm{CO}$ is the mass-to-light ratio of H$_2$ mass to CO emission over a large part of a galaxy. In this definition cloud substructure and even, to some degree, cloud populations are averaged over. Within an individual cloud, the relationship between CO and H$_2$ can be more complex, especially at low metallicity where a large envelope of CO-poor H$_2$ may exist. We will consider three scales for $\alpha_\mathrm{CO}$: the scale of CO-bright clumps, the scale of whole individual complexes, and the whole galaxy. At the small scales of clumps we disregard the H$_2$-rich but CO-poor envelopes of molecular clouds. At the scale of individual atomic-molecular complexes we account for all gas (including CO-poor H$_2$) but results may reflect the local environment or evolutionary state of an individual region. At the scale of the whole galaxy we somewhat marginalize over these conditions.

\subsection{\texorpdfstring{H\,{\footnotesize I}}{HI} Opacity Correction}
\label{subsec:opacity}

Throughout the paper we work with the \hi\ emission without opacity correction. Galaxy-wide studies conclude that local opacity corrections to the column density can exceed an order of magnitude and add globally $20-30$\% to the atomic gas mass \citep[see][and references therein]{Braun09, Kalberla09, Bolatto13}, but without providing clear quantitative prescriptions how to correct observed \mbox{21-cm} \hi\ data sets for optical depth effects. Small-scale or pencil-beam studies within the Milky Way suggest the cold neutral medium (that causes the absorption) to be in compact clouds of parsec-size or narrow filaments and sheets with up to $10{-}100$ pc length \citep{Heiles03, Kalberla09}, but here it remains unknown how these findings extend to galactic scales. Recently, \citet{Bihr15} presented work on the massive cloud complex W43 and advocated for opacity corrections as high as ${\sim} 2.4$ over ${\sim} 100$ pc scales but the mass and surface density of W43 is a factor $\gtrsim 5$ higher than the cloud complexes studied in NGC 6822. Overall, these results highlight that optical depth effects are present in \hi\ observations but also show that we lack a conclusive understanding how to correct \mbox{21-cm} \hi\ observations. Therefore, we adopt the standard assumption of optically thin \hi\ emission and work without opacity correction.

\begin{figure*}[t]
\epsscale{1.15}
\plotone{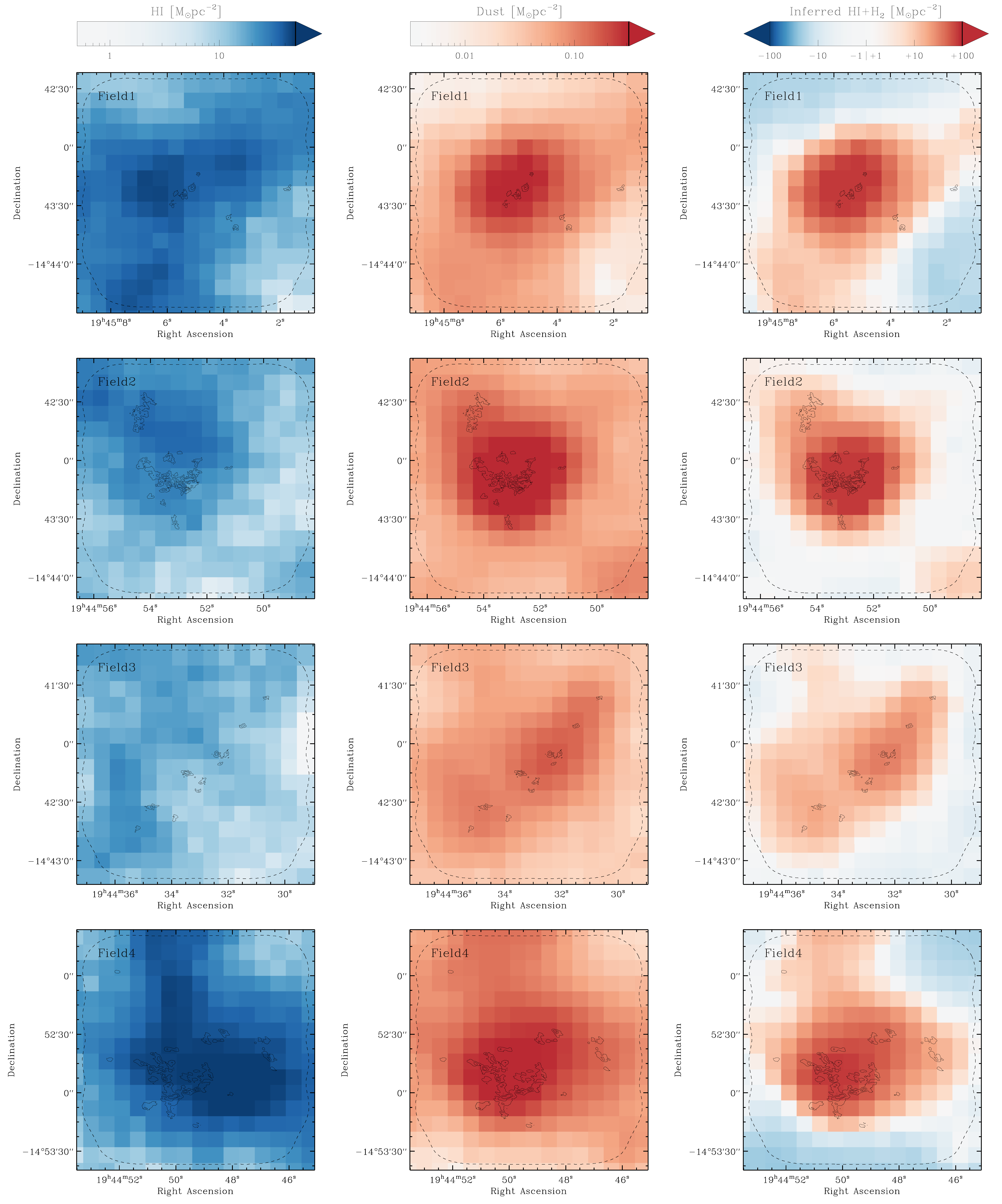}
\caption{High resolution ALMA \cotwo\ data for Fields $1{-}4$ (from top to bottom) shown as contours (at $I_\mathrm{CO} = 2,~10,~20$ \Kkmpers) over low-resolution ($25\arcsec \approx 57$\,pc) grayscale maps of atomic gas (\hi), dust mass from \emph{Herschel} 350~\micron\ data, and dust-inferred (\hi+H$_2$) cloud mass (see text; from left to right) in units of projected mass surface density. The estimate of the latter quantity can lead to negative values locally, in particular at the edges of the survey fields where the gas-to-dust mass ratio is calibrated such that on average no excess emission is found (see text). Therefore, we have adopted a color scale that shows both positive and negative surface densities on a log-stretch.
\label{f5}}
\end{figure*}

\section{Results}
\label{sec:results}

We consider CO emission and molecular gas in NGC 6822 moving from large to small scales. First we derive an estimate of the galaxy-wide CO flux for NGC 6822. Then we consider the structure of the atomic-molecular star-forming complexes that fill our survey fields. We then analyze the local correspondence of CO emission to tracers of the ISM and recent star formation and study the distribution of CO intensities in our survey fields. Finally, we characterize the compact structures seen in our maps, comparing them to similar structures measured in our Galaxy and WLM.

\subsection{Total CO Luminosity of NGC 6822}
\label{subsec:global}

The whole area of NGC 6822 has not yet been mapped in CO, but the galaxy-integrated CO luminosity is important to compare the galaxy to other systems. We estimate this quantity via an ``aperture correction'' from our observed fields to the whole galaxy. To do this, we consider several tracers of recent star formation, H$\alpha$, $24~\micron$, $70~\micron$. These tracers should scale linearly with molecular gas, and so CO emission, over large, $\sim$kpc, spatial scales \citep[e.g.,][]{Schruba11,Leroy13b}. We measure CO luminosity in our fields, and then also measure the luminosity of these tracers of recent star formation both within our fields and over the whole galaxy. Our ``aperture correction'' is the ratio of total star formation tracer luminosity of the galaxy to the luminosity inside our fields. Applying this scale factor to the CO emission from our fields, we estimate the total CO luminosity of the galaxy.

The sum of \cotwo\ luminosities inside our four inner ALMA fields is $3.34 \times 10^4$ \Kkmperspc\ and these fields harbor 63\% of the global H$\alpha$ flux and 65\% of the global \emph{Spitzer} $24~\micron$ flux. Scaling our observed CO luminosity by $1/0.64$, we estimate the global \cotwo\ luminosity to be $5.2 \times 10^4$ \Kkmperspc. Because ALMA recovers only $(73 \pm 20)\%$ of the total flux (see \autoref{sec:data}), we scale this further by $1/(0.7\pm0.2)$ to arrive at a best-estimate global \cotwo\ luminosity of NGC 6822 of $\sim 6{-}10 \times 10^4$ \Kkmperspc. This value agrees within the uncertainty with a similarly derived estimate by \citet{Gratier10a}, who scaled from their IRAM \mbox{30-m} map to estimate a global \cotwo\ luminosity of $\sim 8{-}13 \times 10^4$~\Kkmperspc .

To calculate the \coone\ luminosity, we need to further scale by the ratio $R_{21}^{-1}$. We argue by analogy to other systems that $R_{21} \approx 0.7 - 1.0$ but uncertain. Our best-estimate of the global \coone\ luminosity for NGC 6822 is thus $\sim (10 \pm 5) \times 10^4$ \Kkmperspc. 

This integrated luminosity is small, establishing that NGC 6822 resembles other star-forming, low metallicity dwarf galaxies in showing a low amount of CO luminosity compared to its present day star formation rate, stellar mass, and atomic gas mass. For comparison, our estimate for the total CO luminosity of NGC 6822 is comparable to the CO luminosity of one of our individual Galactic comparison clouds, which have CO luminosities $\sim 5{-}20 \times 10^4$ \Kkmperspc .

\subsection{Atomic-Molecular Complexes Hosting the~CO~Clumps}
\label{subsec:complex}

Our four inner fields host ${\sim} 2/3$ of the star formation activity in NGC 6822. The H$\alpha$ and dust morphology, visible in \autoref{f4} \& \ref{f5}, extend for many tens of parsecs in each region. Though measured at much lower resolution ($25\arcsec = 57$~pc), the \hi\ and dust maps show that our observed CO clumps exist inside larger structures of gas and dust. We expect optically thin dust emission to trace the distribution of gas and \hi\ emission to show the dominant atomic gas reservoir, modulo optical depth effects. We use these to estimate the overall mass and atomic-molecular balance in the star-forming complexes.

We use dust, combined with a gas-to-dust ratio, $\delta_\mathrm{GDR}$, to trace the total gas reservoir,
\begin{eqnarray}
\delta_\mathrm{GDR} \Sigma_\mathrm{dust} = \Sigma_\mathrm{atom} + \Sigma_\mathrm{mol}~.
\label{eq:gdr}
\end{eqnarray}
\noindent Because $\Sigma_\mathrm{atom}$ can be measured directly, we estimate the GDR by comparing $\Sigma_\mathrm{atom}$ and $\Sigma_\mathrm{dust}$ in regions where atomic gas makes up most of the gas. We use the outer 50 pc wide ring in each of our regions, assuming based on the lack of CO emission and bright signatures of high mass star formation that the gas in this ring is mostly atomic. \autoref{t2} lists the GDR with scatter for each of our regions. 

On average, we find $\delta_\mathrm{GDR} = 380 \pm 70$ with only moderate variation among the four fields. This is ${\sim} 2.5$ times higher than the Galactic value but only half the value expected for the ${\sim} 1/5$ solar metallicity of NGC 6822 when assuming an inverse linear scaling of GDR and metallicity \citep{RemyRuyer14}. However, the absolute normalization of dust masses estimated from IR SED modeling remains uncertain at a level that could resolve this discrepancy (see \autoref{subsec:hidust}). Our application of the dust map is to trace gas. For that purpose we require only a linear scaling of dust and gas, any normalization issues are controlled by measuring GDR in the local control field.

Before calculating the mass of the complexes, we subtract a local background from the \hi\ and the dust maps. To do this, we measure the mean surface density in the outer 50 pc ring and subtract this value from the whole \hi\ and dust map for the complex. This isolates the star-forming complex as the excess emission over the diffuse interstellar medium. This is a particular concern for dwarf galaxies which (warm atomic) interstellar medium has large spatial extent, large scale height, and high filling factor \citep[e.g.,][]{Bagetakos11}. This also lets us assess if the star-forming complexes have an atomic gas component in excess of the diffuse atomic gas. No similar subtraction appears necessary for the CO emission.

\begin{figure}[tb]
\epsscale{1.05}
\plotone{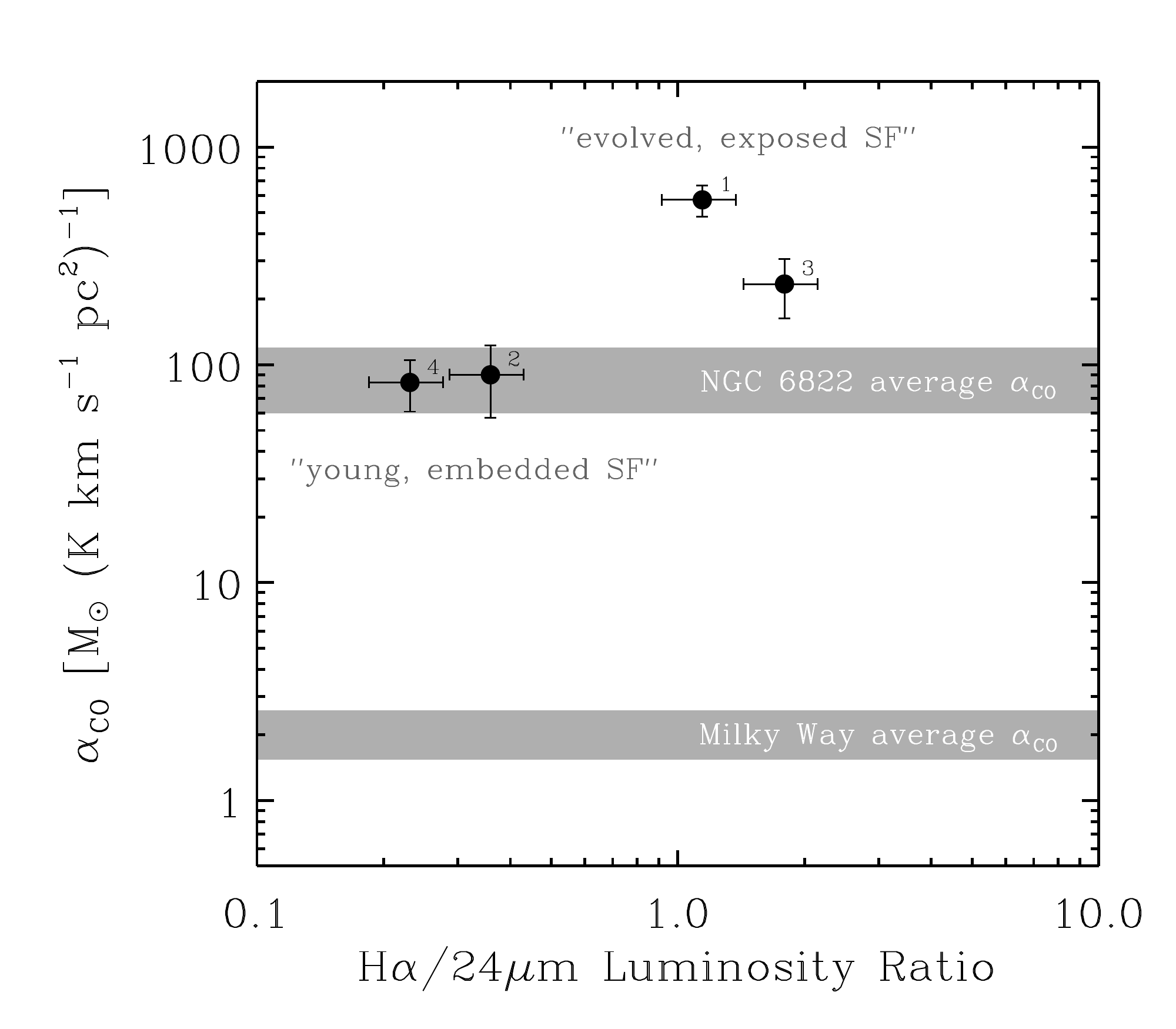}
\caption{CO-to-H$_2$ conversion factor, $\alpha_\mathrm{CO}$, as function of the H$\alpha$-to-24\,$\micron$ luminosity ratio determined over ${\sim} 150$~pc scales for the four atomic-molecular complexes (indexed by field number).
\label{f6}}
\end{figure}

After background subtraction, we find for each region a large concentration of dust (and thus gas) coincident with the star-forming complex (\autoref{f5}). Applying our measured GDR to the dust distribution, we derive total gas masses of $M_\mathrm{gas} = 0.7{-}1.7 \times 10^6$ \Msun\ for each complex. We subtract from the dust-inferred total gas mass the local excess \hi\ mass and find that about $30\%$ of the mass in these complexes is atomic gas traced by the \mbox{21-cm} line. The rest is visible in dust but not in \mbox{21-cm} emission. This is either opaque \hi , H$_2$, or the signature of small-scale variations in the gas-to-dust ratio with the sense of a ${\sim} 3$ times lower gas-to-dust ratio. We proceed interpreting the signal as molecular gas, but note the need for more work in this area \citep[see discussions in][]{Leroy07, Leroy11, Sandstrom13, Jameson16}. In this case, the gas structures hosting star formation in NGC 6822 are about the mass of the biggest Galactic high mass star-forming clouds of order ${\sim} 10^6$~\Msun, with ${\sim} 70\%$ of their mass in the molecular form and ${\sim} 30\%$ of their mass in the atomic form.

The impact of the background subtraction is to remove the diffuse atomic interstellar medium from the excess gas that we associate with the star-forming complexes. This diffuse medium has $\Sigma_\mathrm{atom} = 15 - 25$ \Msunperpc\ in our survey regions and makes ${\sim} 80\%$ of the atomic gas columns along the lines of sight toward the star-forming complexes. In case the reader prefers to associate this diffuse gas to the star-forming complexes, then their total gas masses increase by a factor of ${\sim} 1.6$ and the atomic and molecular gas phases contribute roughly equal amounts to the complexes' mass. We note that the mass of the molecular component of the star-forming complexes --- and thus all results that consider only the molecular gas and its CO emission --- remain unchanged (within $5\%$) by (not) applying the background subtraction.

From the ratio of H$_2$ derived from dust to CO emission, this analysis implies a CO-to-H$_2$ conversion factor. If we carry out a joint solution for $\alpha_\mathrm{CO}$ and GDR across the galaxy \citep[following][]{Sandstrom13, Leroy11}, we derive $\delta_\mathrm{GDR} = 320 \pm 80$ and $\alpha_\mathrm{CO} = 110 \pm 30$ \acounits , about $2$ times and $25$ times the Galactic values, respectively. However, this obscures strong variations among our four fields. \autoref{f6} shows that $\alpha_\mathrm{CO}$ varies systematically with the CO, H$\alpha$, and $24~\micron$ morphology and luminosity. For the two Fields 2 \& 4 with compact CO and H$\alpha$ morphology, dominated by embedded star formation (i.e., low H$\alpha / 24\mu$m ratio), and bright CO emission, we determine $\alpha_\mathrm{CO} = 85 \pm 25$ \acounits , about ${\sim} 20$ times the Galactic value. In the other two Fields 1 \& 3, which have more extended and exposed SFR tracer emission (i.e., high H$\alpha / 24\mu$m ratio) and much lower CO luminosity, $\alpha_\mathrm{CO}$ appears much larger, though uncertain, $\alpha_\mathrm{CO} \approx 235 \pm 72$ and $\approx 572 \pm 93$ \acounits , respectively. We stress that considerable uncertainties remain in this analysis mainly due to the large uncertainty in the dust mass.

We speculate that the large discrepancy in $\alpha_\mathrm{CO}$ between fields is linked to the evolutionary state of the star-forming complex. Early in the history of the complex, the gas is compact and dense. Such structures are good at forming CO and effective at shielding it from dissociating radiation. In this case $\alpha_\mathrm{CO}$ is relatively small (Fields 2~\&~4). Later in a complex's life, the gas gets disrupted by stellar feedback, which is visible as an evolving \hii\ region. With lower density, the ability of gas to form and shield CO is suppressed, driving $\alpha_\mathrm{CO}$ to larger values as more and more H$_2$ survives without CO (Fields 1~\&~3).

\subsection{Coincidence of CO with \texorpdfstring{$8\,\micron$}{8um}, \texorpdfstring{$24\,\micron$}{24um}, and~\texorpdfstring{H$\alpha$}{Ha}~Emission}
\label{subsec:pdrtracers}

\begin{figure}[tb]
\epsscale{1.05}
\plotone{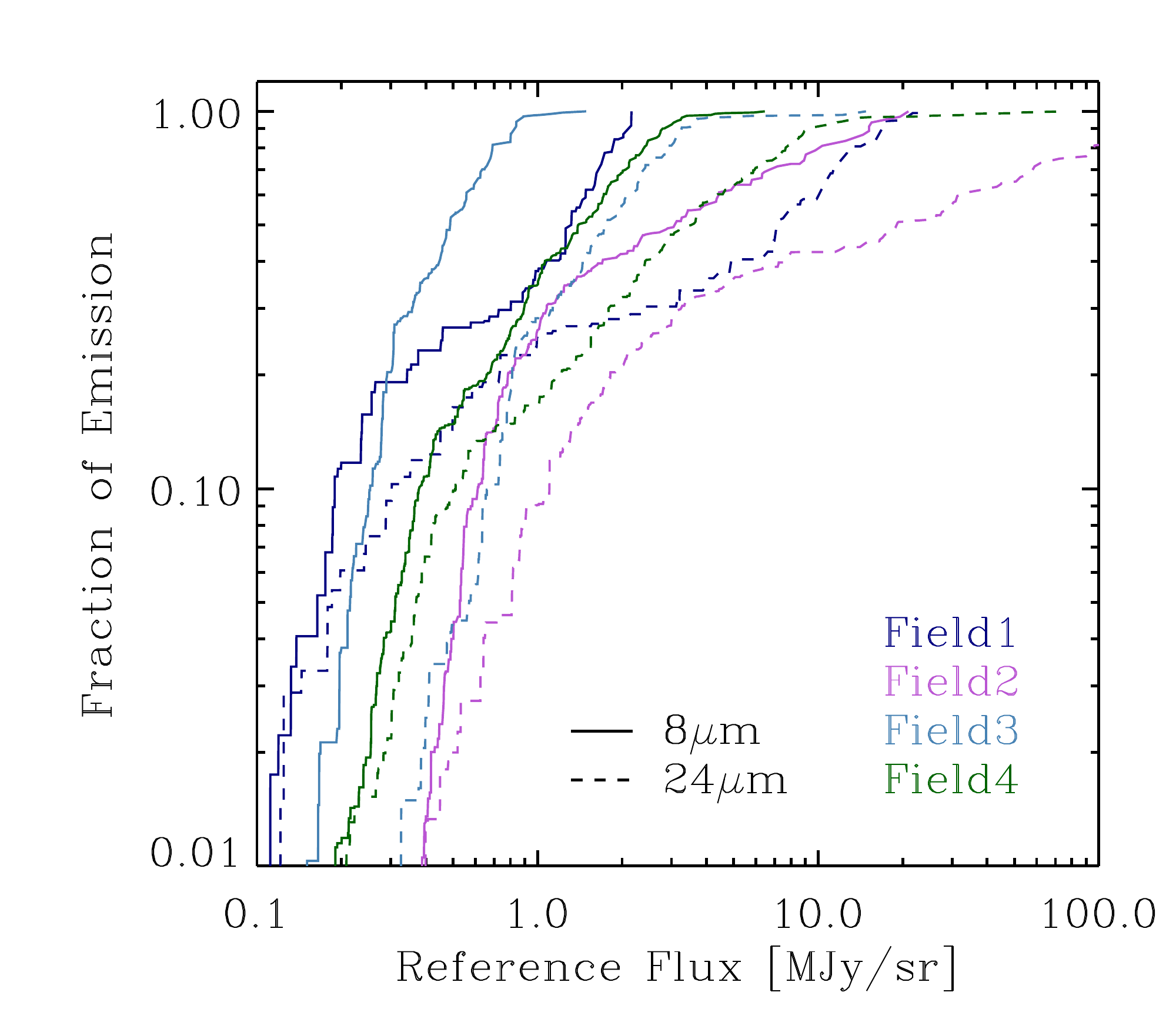}
\caption{Cumulative distribution function of \cotwo\ integrated intensities as function of the $8~\micron$ and $24~\micron$ intensity (solid and dashed lines) at a resolution of $6\arcsec \approx 14$\,pc. CO is more strongly correlated with $8~\micron$ than $24~\micron$ (i.e., steeper rising curves).\label{f7}}
\end{figure}

We targeted our survey toward regions of active star formation, traced by bright H$\alpha$ and mid-IR emission. CO emission tracks these other wavelengths on much larger scales because stars form out of molecular gas. \autoref{f4} shows a more complicated relationship on smaller scales, reflecting the evolution of young stellar populations and the different emission mechanisms at play. In \autoref{t5} we quantify how CO emission correlates with H$\alpha$ and IR emission at resolutions of $2\arcsec \approx 4.6$\,pc (H$\alpha$ and $8~\micron$) and  $6\arcsec \approx 14$\,pc (all three tracers). 

First, we determine the maximum intensity contour for each tracer that encompasses $80\%$ of the total CO emission; that is, we ask what threshold in H$\alpha$ or $8~\micron$ is needed to capture most of the CO. This threshold usually also encompasses a large area with little or no CO emission. To quantify how large this area is, and so the closeness of correspondence with CO, we report an ``area fraction,'' which is the ratio of the area inside the 80\% CO threshold for that tracer to the 80\% threshold for CO. Finally, we report the Spearman rank correlation coefficient between each tracer and CO above this threshold. 

\capstartfalse
\setlength{\tabcolsep}{3pt}
\begin{deluxetable}{lcccc}
\tablecolumns{5}
\tablecaption{Association of the top 80\% of CO and IR Emission\label{t5}}
\tablehead{\colhead{Data} & \colhead{Flux Cut\tablenotemark{a}} & \colhead{Flux Fraction\tablenotemark{b}} & \colhead{Area Fraction\tablenotemark{c}} & \colhead{Rank Corr.}}
\startdata
\cutinhead{Resolution of $2^{\prime\prime} \approx 4.6$\,pc}
CO & $1.6 \pm 0.7$ & $0.79 \pm 0.09$ & 1.0 & 1.0 \\
H$\alpha$ & $0.003 \pm 0.013$ & $0.77 \pm 0.27$ & $23 \pm 22$ & $0.19 \pm 0.19$ \\
8\,$\mu$m & $0.65 \pm 0.21$ & $0.78 \pm 0.14$& $5.1\pm 4.6$ & $0.41 \pm 0.10$  \\[-1mm]
\cutinhead{Resolution of $6^{\prime\prime} \approx 14$\,pc}
CO & $0.5 \pm 0.3$ & $0.81 \pm 0.13$ & 1.0 & 1.0 \\
H$\alpha$ & $0.005 \pm 0.010$ & $0.77 \pm 0.25$ & $6.9 \pm 6.5$ & $0.20 \pm 0.34$ \\
8\,$\mu$m & $0.40 \pm 0.24$ & $0.82 \pm 0.15$ & $3.4 \pm 2.3$ & $0.54 \pm 0.09$ \\
24\,$\mu$m & $1.1 \pm 0.5$ & $0.80 \pm 0.09$ & $3.4 \pm 2.0$ & $0.37 \pm 0.16$
\enddata
\tablenotetext{a}{Common flux threshold for CO\,[K\,km\,s$^{\shortminus 1}$], or H$\alpha$ or IR [MJy\,sr$^{\shortminus 1}$] holding 80\% of the CO flux within each region.}
\tablenotetext{b}{Fraction of CO, H$\alpha$, IR flux in each regions above common threshold.}
\tablenotetext{c}{Fraction of area covered by H$\alpha$ or IR common threshold versus area holding $80\%$ of the CO emission.}
\end{deluxetable}
\setlength{\tabcolsep}{6pt}
\capstarttrue

We find the closest correspondence between CO and $8~\micron$ emission and the weakest relationship between CO and H$\alpha$, with $24~\micron$ intermediate. This reflects the different emission mechanisms at play. The $8~\micron$ emission originates from PAH molecules and traces PDRs. Despite a few compact sources, most emission comes from extended, diffuse structures which coincide with the CO emission. Overall, the distribution of the two tracers matches well. \citet{Gratier10a} noted a similar good correspondence at larger scales.

The $24~\micron$ emission traces warm dust heated by young, embedded stars. At this scale, $24~\micron$ sources correspond to an early, embedded phase of star formation that takes place before feedback can disrupt the parent gas cloud. The distribution of $24~\micron$ intensity is concentrated into a few ($\sim 1{-}2$) bright sources per survey field. Each $24~\micron$ source is associated with a CO-bright structure, but the reverse is not true. Many CO peaks lack a corresponding $24~\micron$ source. 

H$\alpha$ and photospheric UV emission are most visible after the embedded phase traced by $24~\micron$ emission. At the scales that we probe, H$\alpha$ emission is not co-spatial with the CO emission, reflecting both the disruption of clouds and the finite extent of \hii\ regions. Instead, we see the neutral gas (including the CO-bright clumps) swept up at the boundary of the \hii\ region bubbles --- a picture that is well known from Galactic star-forming clouds \citep[e.g., the W3/W4 molecular cloud--\hii\ region complex; see images by][]{Bieging11}.

Thus the $8~\micron$ emission correlates most directly with CO emission. The areal extent of the $8~\micron$ contour needed to capture 80\% of the CO emission is the smallest of the tracers that we test but the correspondence is not perfect. This $8~\micron$ contour is still a factor of ${\sim} 4$ larger than the actual CO-bright regions. The $24~\micron$ shows the second best correlation over similar areal coverage, but the scaling between CO and $24~\micron$ emission on these scales is strongly nonlinear. H$\alpha$ shows only marginal correlation and the encompassing H$\alpha$ contour has large areal extent. 

\autoref{f7} shows this result by plotting the cumulative distribution of CO integrated intensity as function of the threshold intensity at $8~\micron$ or $24~\micron$ at 2~pc resolution. The curves for $8~\micron$ are steep, and high CO fractions are reached over a small intensity range near $I_\mathrm{8 \mu m} \approx 0.5$~MJy~sr$^{-1}$. A contour with this intensity does a good job of predicting where bright CO emission may be found. Our results reinforce the idea from \citet{Sandstrom10} and \citet{Gratier10a} that PAH emission can provide a predictor of the location, and perhaps strength, of CO emission in low metallicity environments. We have used $8~\micron$ here; given the full-sky coverage of the Wide-field Infrared Survey Explorer (WISE) it will be important to test whether the same conclusions would apply using the $12~\micron$ PAH feature covered by WISE's band~3.

\subsection{Distribution of CO Intensities}
\label{subsec:pixelwise}

\begin{figure}[tb]
\epsscale{1.05}
\plotone{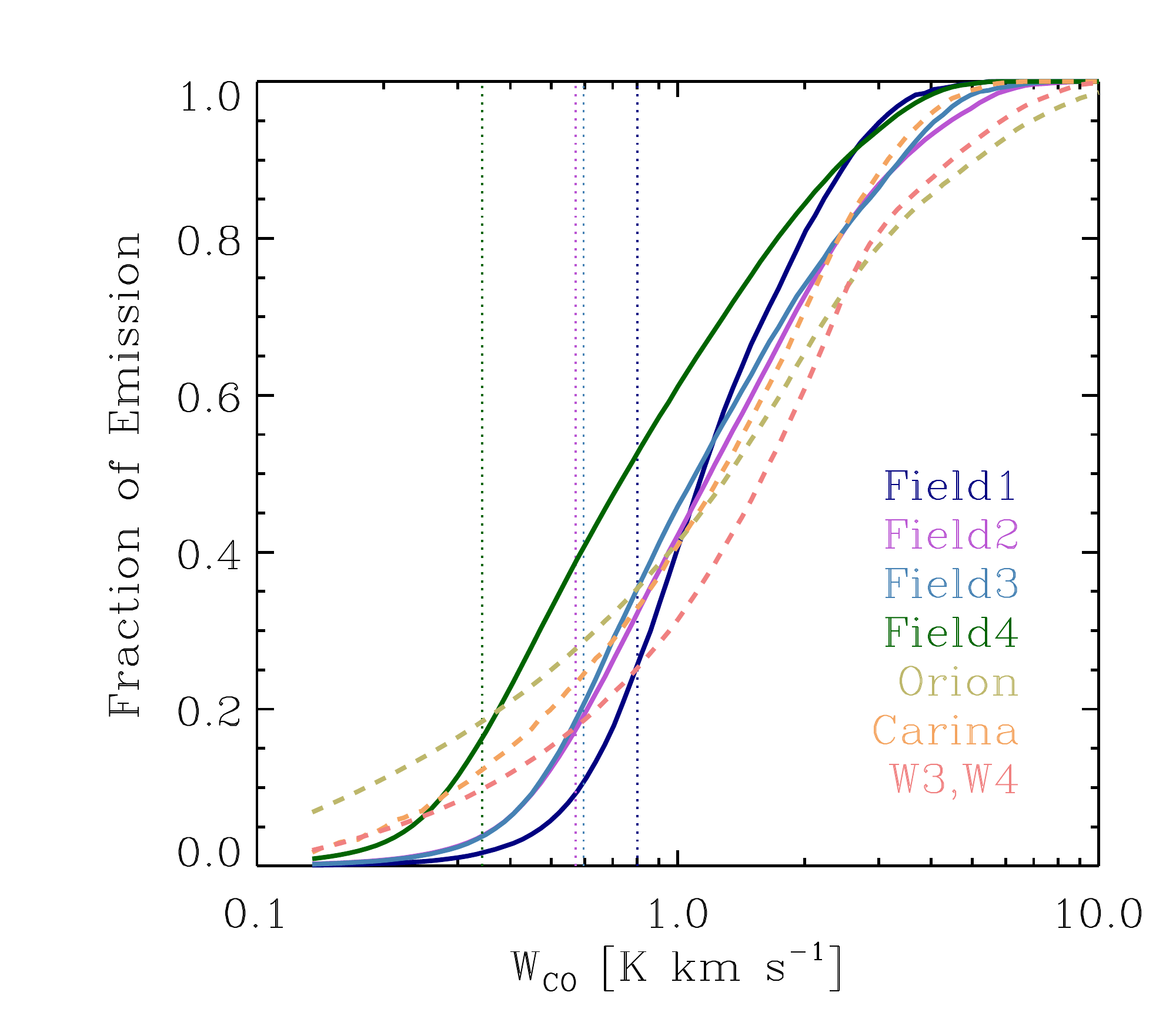}
\caption{Fraction of total CO emission above a varying CO pixel intensity threshold for NGC 6822 (solid lines) and a small reference sample of matched-resolution \coone\ or \cotwo\ data from Galactic molecular clouds of similar mass and SFR (dashed lines; see text). No scaling between \coone\ and \cotwo\ intensities has been applied. The vertical dotted lines show two times the rms sensitivities for the NGC 6822 survey fields; the distributions may be incomplete (i.e., lower limits) below the dotted lines due to signal masking and missing extended emission in our ALMA data.
\label{f8}}
\end{figure}

The distribution of intensities in our survey provides a basic measure of the structure of the CO-emitting gas in NGC 6822. We show this in \autoref{f8}, where we plot the fraction of total flux above a specified CO pixel intensity threshold (expressed as intensity times channel width, $W_\mathrm{CO}$). For reference we also plot the distributions for our three Galactic comparison clouds after convolving them to the resolution of our data (2\,pc $\times$ 0.635\,\kmpers ) and applying the same masking procedure that we use for NGC 6822. 

The distributions in NGC 6822 differ from those of the Milky Way clouds, with emission coming from a narrower range of intensities in the NGC 6822 data. Less flux and less pixels show either low or high intensities. The absence of low intensity emission may be explained physically, corresponding to a suppression of CO abundance in regions that are weakly shielded. However, given uncertainties in flux recovery and the limited sensitivity of our data, this difference is not significant. The difference at higher intensity does appear significant but modest in strength: on average, the NGC 6822 cdfs are shifted to ${\sim} 30\%$ lower pixel intensities with less flux in the brightest regions than the Orion or Carina molecular clouds.

\subsection{CO Clump Properties}
\label{subsec:clumpwise}

\capstartfalse
\setlength{\tabcolsep}{3pt}
\begin{deluxetable*}{r*{14}{r}}
\tablecolumns{14}
\tablecaption{Molecular Cloud Catalog\label{t6}}
\tablehead{\colhead{ID} & \colhead{R.A.} & \colhead{Decl.} & \colhead{$V_{\rm LSR}$} & \colhead{$T_{\rm pk}$} & \colhead{$S/N$} & \colhead{$R_{\rm maj}$} & \colhead{$R_{\rm min}$} & \colhead{PA} & \colhead{$R$} & \colhead{$\sigma_{\rm v}$} & \colhead{$L_{\rm CO}$} & \colhead{$M_{\rm vir}$} & \colhead{$\alpha_{\rm vir}$} \\
\colhead{} & \colhead{(hh:mm:ss.s)} & \colhead{(dd:mm:ss)} & \colhead{(km\,s$^{\shortminus 1}$)} & \colhead{(K)} & \colhead{} & \colhead{(pc)} & \colhead{(pc)} & \colhead{(deg)} & \colhead{(pc)} & \colhead{(km\,s$^{\shortminus 1}$)} & \colhead{(caption)} & \colhead{($10^2$\,\Msun)} & \colhead{} \\
\colhead{(1)} & \colhead{(2)} & \colhead{(3)} & \colhead{(4)} & \colhead{(5)} & \colhead{(6)} & \colhead{(7)} & \colhead{(8)} & \colhead{(9)} & \colhead{(10)} & \colhead{(11)} & \colhead{(12)} & \colhead{(13)} & \colhead{(14)}}
\startdata
   1 &  19:45:03.8 &  -14:43:36 & -45.0 &   1.5 &   2.4 &   0.9 &   0.3 &    96 &   0.5 $\pm$   0.1 &  0.19 $\pm$  0.00 &   0.0 $\pm$   0.0 &    0.2 $\pm$    0.0 &   1.3 $\pm$   0.2 \\
   2 &  19:45:03.6 &  -14:43:40 & -44.0 &   3.3 &   5.2 &   3.3 &   2.5 &    62 &   2.9 $\pm$   0.6 &  0.58 $\pm$  0.12 &   0.5 $\pm$   0.2 &   10.1 $\pm$    3.8 &   5.0 $\pm$   2.8 \\
   3 &  19:45:01.7 &  -14:43:21 & -42.0 &   2.4 &   3.8 &   2.9 &   1.8 &   148 &   2.3 $\pm$   0.9 &  0.46 $\pm$  0.02 &   0.2 $\pm$   0.0 &    5.1 $\pm$    2.1 &   6.9 $\pm$   2.9 \\
   4 &  19:45:03.8 &  -14:43:36 & -40.0 &   4.2 &   6.6 &   3.6 &   2.7 &   131 &   3.1 $\pm$   0.8 &  1.26 $\pm$  0.02 &   1.2 $\pm$   0.0 &   51.9 $\pm$   13.9 &   9.6 $\pm$   2.6 \\
   5 &  19:45:01.7 &  -14:43:20 & -40.0 &   2.3 &   3.7 &   2.3 &   1.2 &     6 &   1.7 $\pm$   0.4 &  0.39 $\pm$  0.05 &   0.1 $\pm$   0.0 &    2.6 $\pm$    0.9 &   4.5 $\pm$   1.7 \\
   6 &  19:45:03.6 &  -14:43:41 & -39.0 &   5.4 &   8.5 &   3.7 &   1.9 &   144 &   2.6 $\pm$   0.5 &  2.06 $\pm$  0.28 &   1.8 $\pm$   0.5 &  116.9 $\pm$   30.9 &  14.5 $\pm$   5.7 \\
   7 &  19:45:01.9 &  -14:43:21 & -39.0 &   2.1 &   3.4 & \nodata & \nodata & \nodata &   1.1 $\pm$     0 &  0.36 $\pm$  0.04 &   0.1 $\pm$   0.0 &    1.5 $\pm$   0.0 &   3.6 $\pm$   0.0 \\
   8 &  19:45:01.7 &  -14:43:21 & -37.0 &   4.6 &   7.4 &   4.4 &   1.1 &     2 &   2.2 $\pm$   0.5 &  1.10 $\pm$  0.11 &   1.4 $\pm$   0.5 &   27.4 $\pm$    7.4 &   4.3 $\pm$   1.9 \\
   9 &  19:45:01.7 &  -14:43:19 & -36.0 &   3.5 &   5.5 & \nodata & \nodata & \nodata &   0.5 $\pm$     0 &  0.53 $\pm$  0.08 &   0.3 $\pm$   0.1 &    1.4 $\pm$   0.0 &   1.2 $\pm$   0.0 \\
  10 &  19:45:05.1 &  -14:43:20 & -35.0 &   2.1 &   3.3 &   5.8 &   1.2 &   176 &   2.6 $\pm$   0.2 &  0.45 $\pm$  0.01 &   0.2 $\pm$   0.0 &    5.6 $\pm$    0.6 &   8.3 $\pm$   0.8 \\
\nodata & \nodata & \nodata & \nodata & \nodata & \nodata & \nodata & \nodata & \nodata & \nodata & \nodata & \nodata & \nodata & \nodata \\
\nodata & \nodata & \nodata & \nodata & \nodata & \nodata & \nodata & \nodata & \nodata & \nodata & \nodata & \nodata & \nodata & \nodata \\
\nodata & \nodata & \nodata & \nodata & \nodata & \nodata & \nodata & \nodata & \nodata & \nodata & \nodata & \nodata & \nodata & \nodata \\
 156 &  19:44:49.0 &  -14:52:55 & -41.0 &   2.7 &  10.0 &   7.5 &   2.2 &     6 &   4.1 $\pm$   0.3 &  1.63 $\pm$  0.11 &   2.2 $\pm$   0.0 &  113.4 $\pm$   13.7 &  12.0 $\pm$   1.5
\enddata
\tablecomments{(1) Cloud identification number (ID); (2) right ascension (R.A.\ (J2000)); (3) declination (Decl.\ (J2000)); (4) velocity ($V_{\rm LSR}$); (5) peak brightness temperature ($T_{\rm pk}$); (6) peak signal-to-noise ratio ($S/N$); (7) semi-major axis length ($R_{\rm maj}$); (8) semi-minor axis length ($R_{\rm min}$); (9) position angle of cloud major axis, measured from R.A.\ through Decl.\ (PA); (10) radius ($R$); (11) velocity dispersion ($\sigma_{\rm v}$); (12) CO luminosity ($L_{\rm CO}$ [$10^2$\,K\,km\,s$^{\shortminus 1}$\,pc$^2$]); (13) virial mass ($M_{\rm vir}$); (14) virial parameter ($\alpha_{\rm vir}$). \\ (This table is available in its entirety in a machine-readable form in the online journal. A portion is shown here for guidance.)}
\end{deluxetable*}
\setlength{\tabcolsep}{6pt} 
\capstarttrue

\capstartfalse
\begin{deluxetable*}{ll*{13}{r}}
\tablecolumns{15}
\tablecaption{Average Properties of Molecular Clouds\label{t7}}
\tablehead{\multicolumn{2}{c}{Target} & \colhead{WLM} & \colhead{} & \multicolumn{5}{c}{NGC 6822} & \colhead{} & \multicolumn{5}{c}{Outer Milky Way} \\[0.5mm]
\cline{3-3} \cline{5-9} \cline{11-15} \\[-1mm]
\colhead{Property} & \colhead{Unit} & \colhead{Median} & \colhead{} & \colhead{Min} & \colhead{$25th$} & \colhead{Median} & \colhead{$75th$} & \colhead{Max} & \colhead{} & \colhead{Min} & \colhead{$25th$} & \colhead{Median} & \colhead{$75th$} & \colhead{Max}}
\startdata
$R$ & pc &     2.21 &  &     0.27 &     1.64 &     2.29 &     3.14 &     7.23 &  &     0.50 &     1.25 &     1.87 &     2.79 &    14.87 \\
$\sigma_\mathrm{v}$ & km s$^{-1}$ &     0.77 &  &     0.17 &     0.68 &     1.06 &     1.40 &     2.84 &  &     0.30 &     0.60 &     0.82 &     1.14 &     8.47 \\
$\sigma_\mathrm{v}^2/R$ & km$^2$ s$^{-2}$ pc$^{-1}$ &     0.35 &  &     0.01 &     0.26 &     0.45 &     0.82 &     2.62 &  &     0.03 &     0.21 &     0.36 &     0.65 &    12.16 \\
$L_\mathrm{CO}$ & $10^2$\,K\,km\,s$^{\shortminus 1}$\,pc$^2$ &     0.69 &  &     0.02 &     0.40 &     1.13 &     2.42 &     8.05 &  &     0.04 &     0.41 &     0.90 &     2.34 &   180.05 \\
$M_\mathrm{lum}$ & $10^2$ \Msun &     2.99 &  &     0.09 &     1.75 &     4.90 &    10.54 &    35.03 &  &     0.15 &     1.67 &     3.70 &     9.61 &   738.22 \\
$M_\mathrm{vir}$ & $10^2$ \Msun &    13.91 &  &     0.18 &     8.52 &    27.41 &    63.13 &   530.69 &  &     0.57 &     5.40 &    12.90 &    31.83 &  2096.26 \\
$\alpha_\mathrm{vir}$ &  &     4.38 &  &     0.46 &     3.22 &     5.00 &     8.27 &    83.07 &  &     0.44 &     1.97 &     3.11 &     5.32 &    61.25 \\
$\mathcal{M}$ &  &     3.92 &  &     0.89 &     3.48 &     5.39 &     7.14 &    14.46 &  &     1.54 &     3.07 &     4.20 &     5.80 &    43.13 \\
$\Sigma_\mathrm{\scriptscriptstyle{FWHM}}$ & \Msun\ pc$^{-2}$ &    17.59 &  &     1.02 &    11.46 &    22.97 &    40.95 &   302.62 &  &     4.69 &    18.81 &    27.27 &    42.28 &   382.70 \\
$n_\mathrm{\scriptscriptstyle{FWHM}}$ & $10^2$ cm$^{-3}$ &     1.47 &  &     0.12 &     1.40 &     3.63 &     7.46 &   156.09 &  &     0.18 &     1.56 &     2.79 &     5.04 &    44.01 \\
$\tau_\mathrm{ff,\scriptscriptstyle{FWHM}}$ & Myr &     2.83 &  &     0.35 &     1.61 &     2.31 &     3.73 &    12.75 &  &     0.38 &     1.39 &     1.87 &     2.50 &     9.51 \\
$\tau_\mathrm{cross,\scriptscriptstyle{FWHM}}$ & Myr &     2.22 &  &     0.41 &     1.15 &     1.64 &     2.21 &     4.61 &  &     0.21 &     1.07 &     1.57 &     2.31 &    12.29
\enddata
\tablenotetext{a}{All values are extrapolated to zero-flux. Assuming $R_{21} = 1.0$ and $\alpha_{\rm CO} = 4.35$ \acounits.}
\end{deluxetable*}
\capstarttrue

\begin{figure*}[t]
\epsscale{1.15}
\plotone{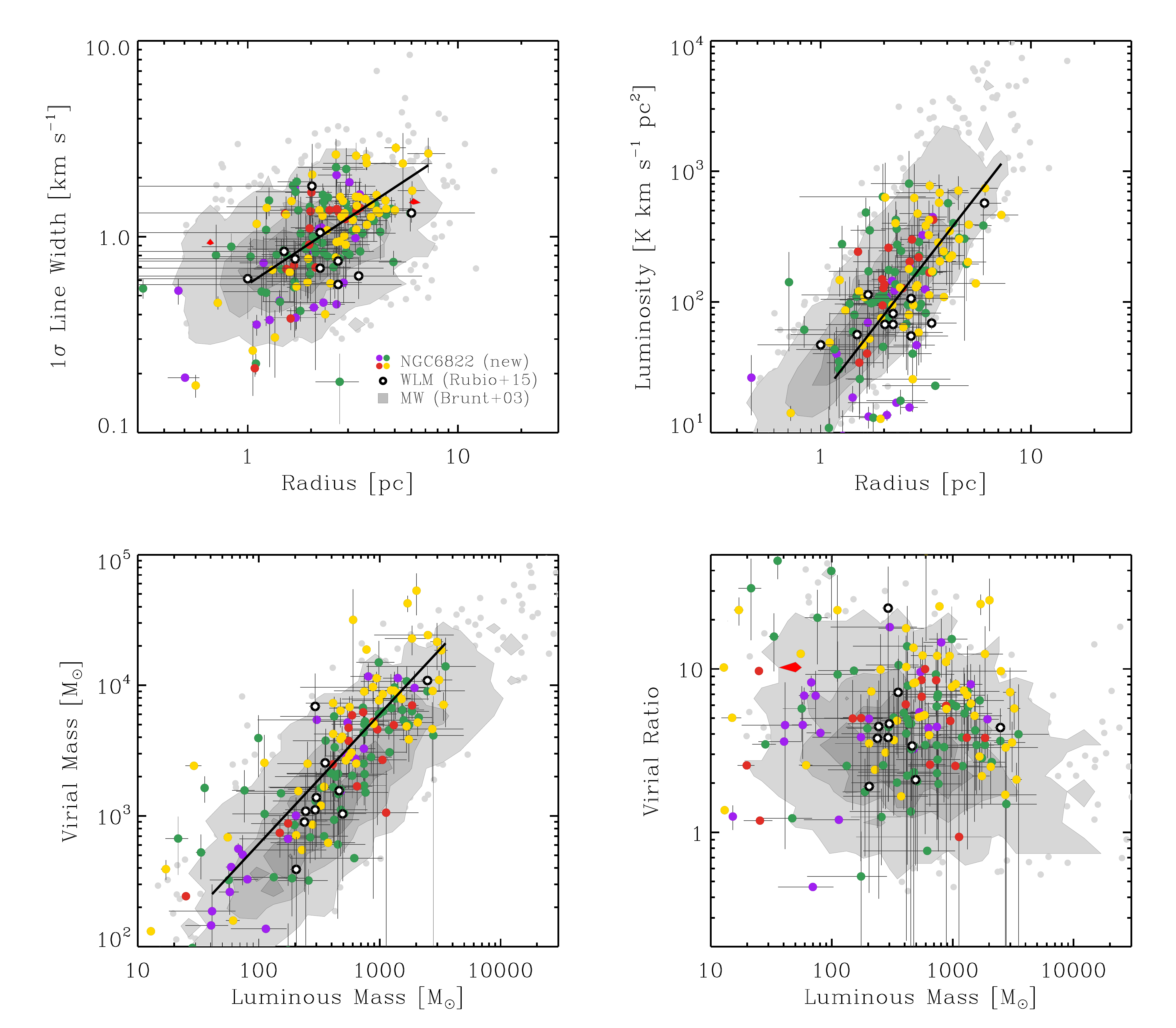}
\caption{Scaling relations for molecular clouds in NGC 6822 of the various survey fields (colored symbols), in WLM \citep[][open black symbols]{Rubio15}, and the outer Milky Way \citep[][gray contours of data density at 5, 35, 65, 95\% completeness and individual gray points in regions of low data density]{Brunt03}. The black lines show power law fits to the NGC 6822 data. The ALMA \cotwo\ data for NGC 6822 are scaled by $R_{21} = 1$ and $\alpha_\mathrm{CO} = 4.35$ \acounits\ to get luminosities and masses. We find good agreement between the cloud properties of NGC 6822, WLM, and the outer Milky Way with offsets $\lesssim 2$ that can be explained by either CO excitation ($R_{21}$), the CO-to-H$_2$ conversion factor ($\alpha_\mathrm{CO}$), or external pressure ($P_\mathrm{ext}$).
\label{f9}}
\end{figure*}

Our survey recovers a number of compact CO structures. We identify and measure the properties of ${\sim} 150$ of these roughly pc-scale ``clumps,'' estimating a size, line width, and CO luminosity for each. We use combinations of these properties to assess the surface and volume density, strength of turbulence, and dynamical state of the clumps. \autoref{t6} lists the inferred properties of each clump while \autoref{t7} lists their average values and dynamic range. We compare these to results for similarly sized structures in WLM \citep{Rubio15} and the outer Milky Way \citep{Brunt03}; their average values and dynamic range\footnote{For WLM we only state the median due to small sample size.} are also listed in \autoref{t7}. \autoref{f9} shows these comparisons, with the WLM data shown as open black dots and the Galactic clumps shown as grayscale contours of data density. Black lines show power law fits to the NGC 6822 clumps, which are useful for comparison to the Galactic distribution. 

The upper left panel of \autoref{f9} shows the line width of a clump as a function of its size. For clumps in virial equilibrium, the amplitude of turbulence at fixed size reflects the surface density of the structure. Regardless of dynamical state, one might interpret a higher dispersion at fixed size as stronger turbulence --- modulo temperature effects this will correspond to a higher Mach number. Differences among the three populations are small in this parameter space, but there is some tendency for NGC 6822 clumps to have higher line width at ${\sim} 5$~pc sizes.

Beyond the scaling, the absolute values shown in the top left panel of \autoref{f9} bear comment. The bright CO-emitting structures in NGC 6822 are remarkably small, a few pc across, with typical rms line widths ${\sim} 1$ \kmpers . This highlights a fundamental result for CO in dwarf galaxies, seen by \citet{Rubio15} and extended here to 150 objects: CO emission comes mostly from compact, narrow line width structures. This observation is only possible with the high resolution and sensitivity of ALMA, so that to our knowledge this is the first time that a large set of clump properties at 2~pc spatial scale have been measured for an external galaxy.

The upper right panel shows the CO luminosity as a function of clump size. As shown in the previous section, clumps in NGC 6822 have average CO surface brightness ${\sim} 30\%$ lower than those in the Milky Way. This is within the systematic uncertainties on flux recovery and excitation, but could also indicate suppressed CO emission in NGC 6822 or that the clumps are not fully resolved. Again, agreement among clump properties for the three galaxies is more notable than differences given how under-luminous NGC 6822 appears in its ratio of global CO luminosity to other quantities.

The bottom left panel shows the virial mass, the mass implied by the clump's size and line width when assuming a simple geometry and virial equilibrium ($M_\mathrm{vir}~[\Msun] = 1040 R \sigma^2~[\mathrm{pc}\,\mathrm{km}^2\,\mathrm{s}^{-2}]$), as a function of clump mass estimated from the CO luminosity using a fiducial Galactic CO-to-H$_2$ conversion factor $\alpha_\mathrm{CO} = 4.35$ \acounits\ and $R_{21} = 1$. There is a good correspondence between clump virial mass and clump luminosity within each sample, but the NGC 6822 clumps show higher virial masses at a given CO luminosity than Milky Way clumps. This offset is not large, less than a factor of ${\sim} 2$, but appears significant.

The bottom right panel shows similar information. We plot the virial parameter, the ratio of virial to luminous mass, as a function of the luminous mass. Assuming that the conversion factor adopted to estimate the luminous mass is correct, then clumps with virial ratio of ${\sim} 1$ are in virial equilibrium, while clumps with virial ratios of ${\sim} 2$ are marginally bound. Those with higher ratios are either pressure-confined or gravitationally unbound. This plot shows clumps in the outer Galaxy to be marginally bound, with $\alpha_\mathrm{vir} \approx 2.5$, while clumps in NGC 6822 are offset by a factor of ${\sim} 2$ toward larger virial parameters. This has two natural interpretations. Either the use of a Galactic conversion factor leads us to underestimate the clump mass in NGC 6822 or the clumps in this galaxy are unbound or pressure-confined.

The main result from \autoref{f9} is that the properties of clumps in the three galaxies are similar independently of their host galaxy's mass or metallicity. Differences are small, a factor of $\lesssim 2$, and comparable to what we might expect from differing methodologies, flux recovery issues, assumptions about excitation. Still, there is evidence here for either a different, more pressure-confined dynamical state for the NGC 6822 clumps or a slightly higher CO-to-H$_2$ conversion factor on scales of a few parsecs. We return to the implications of these results in \autoref{subsec:stability} \& \ref{subsec:sfgas}.

\section{Discussion}
\label{sec:discussion}

\subsection{The CO-bright Clumps}
\label{subsec:clumps}

We identify ${\sim} 150$ compact CO structures in our cubes. These CO-bright clumps have an average size of $R \approx 2.3$ pc, velocity dispersion of $\sigma_\mathrm{v} \approx 1.1$ \kmpers, and virial mass of $M_\mathrm{vir} \approx 2.7 \times 10^3$ \Msun\ (see \autoref{t7}). These sizes and masses are small compared to a whole giant molecular cloud, so we have referred to these as ``clumps.'' The very small size of these clumps is a main observational result of this paper. It shows that many previous results regarding the faintness of CO in dwarf galaxies can be explained in terms of beam filling: bright CO is confined to small structures that occupy only a small portion of the available area. Observations with physical beams larger than our 2~pc will dilute this intensity and so will recover faint CO emission.

Following an analysis presented by \citet{Rubio15} for WLM, we evaluate the dynamical state of the clumps in three ways: (a) Assuming approximate virialization, so that $M_\mathrm{vir} \approx M_\mathrm{gas}$, the clumps have a gas mass surface density of $\Sigma_\mathrm{gas} = 125$ \Msunperpc\ and an H$_2$ volume density of $n_\mathrm{H_2} = 1000$ cm$^{-3}$. (b) We can consider pressure equilibrium between the internal turbulent pressure of the CO clumps and external pressure by the weight of the overlaying gas layers of the atomic-molecular complex and the diffuse atomic medium. The external pressure due to self-gravity on the CO clump is $P_\mathrm{ext} / k_\mathrm{B} = \pi/2 G \Sigma_\mathrm{gas}^2 = 5 \times 10^5$ \mbox{K\,cm$^{-3}$} where $\Sigma_\mathrm{gas} = 120$ \Msunperpc\ is the sum of gas columns of the atomic-molecular complex and the diffuse ISM. This pressure has to be balanced by the internal pressure $P_\mathrm{int} / k_\mathrm{B} = \rho \sigma_\mathrm{v}^2$ due to turbulence which, if turned around, requires an H$_2$ density of the clumps of $n_\mathrm{H_2} \approx 1200$ cm$^{-3}$. (c) The volume density required for collisional excitation of CO rotational transitions is $n_\mathrm{H_2} \approx 1000$ cm$^{-3}$ \citep{Glover12b}. Overall, we conclude that the virial density, excitation density, and pressure equilibrium density for the H$_2$ gas in the CO-bright clumps are quite similar around $n_\mathrm{H_2} \approx 1000$ cm$^{-3}$ and therefore considering these gas structures to be in (quasi) dynamical equilibrium seems justified.

\subsection{The Comparison to Milky Way Clumps}
\label{subsec:comparison}

Our analysis of the CO-bright clumps in NGC 6822 has shown that their macroscopic properties are very similar (with possible offsets less than a factor of two) to CO-bright structures within our Galaxy. For our adopted $R_{21} = 1$ and $\alpha_\mathrm{CO} = 4.35$ \acounits, the clumps in NGC 6822 have ${\sim} 30\%$ lower surface brightness and a factor of $\lesssim 2$ higher virial masses or virial ratios for fixed (CO-inferred) luminous masses. There are three easy explanations which can explain the (apparent) small difference to the Galactic observations. (1) The CO excitation is sub-thermal and $R_{21}$ is closer to the standard value ($\sim 0.7$) derived for Milky Way clouds or massive disk galaxies. This would directly remove the difference in surface brightness but leave clumps in NGC 6822 to be slightly less bounded than in the Milky Way. (2) The CO-to-H$_2$ conversion factor is larger (by a factor of $\sim 2$) than the standard value for Milky Way clouds or massive disk galaxies, which would be motivated by the lower surface brightness (due to a lower beam filling factor) and results in identical virial ratios in NGC 6822 and the Milky Way. (3) The more massive shielding layers around the CO-bright clumps in NGC 6822 provide sufficient external pressure to bring the internal turbulent pressure and self-gravity into a dynamical balance. In the previous sections we have seen that the required changes listed here are all consistent with our measurements within the uncertainties and thus either of the three cases or a combination of them provide a viable pathway to explain our observations.

We can therefore conclude that the structure of the atomic-molecular complexes and the CO-bright clumps in NGC 6822 seems normal in terms of density, pressure, and column density, and that they appear to be marginally gravitationally bound. The main effect of the lower CO abundance in NGC 6822 is to lower the brightness and extent of the CO-bright structures. This explains why the clumps in NGC 6822 and the Milky Way populate the same loci in plots of scaling relations of their macroscopic properties (see \autoref{subsec:clumpwise}). \citet{Rubio15} have arrived at the same set of conclusions in their analysis of ten CO-bright clumps in WLM which has only ${\sim} 1/8$ solar metallicity. Taken these results together, we conclude that clumps inside molecular clouds appear to be in virial equilibrium after the non-negligible contribution of surface pressure from the surrounding gas is accounted for. We find no dependence on gas phase metallicity over ${\sim} 0.2 - 1$ solar metallicity. The dynamical state of dense clumps --- the location where stars form --- is thus indistinguishable between clouds in the Milky Way or nearby low metallicity dwarf galaxies.

\subsection{The CO-to-\texorpdfstring{H$_\mathit{2}$}{H2} Conversion Factor}
\label{subsec:aco}

We have determined the CO-to-H$_2$ conversion factor, $\alpha_\mathrm{CO}$, on several scales. On the largest scales, our result mirrors the discussion in \autoref{subsec:comparison}: integrating over our dust-inferred molecular gas masses and dividing by the ALMA \cotwo\ luminosity, we find $\alpha_\mathrm{CO(2-1)} \approx 110$~\acounits . Adjusting this to account for our expectation of ${\sim} 73\%$ flux recovery lowers this number to ${\sim} 80$ \acounits . For a \coone\ conversion factor, this should be modified by the appropriate line ratio; we argue that a value close to thermal (${\sim} 1$ in K units) is likely appropriate for a dwarf irregular galaxy like NGC 6822. This large-scale $\alpha_\mathrm{CO}$ value is ${\sim} 20{-}25$ times larger than the Milky Way value of $\alpha_\mathrm{CO} = 4.35$~\acounits . Our $\alpha_\mathrm{CO}$ estimate is a factor of ${\sim} 4.7$ larger than the calculations by \citet{Leroy11}, who used {\em Spitzer} and IRAM \mbox{30-m} data by \citet{Gratier10a} to derive $\alpha_\mathrm{CO(2-1)} = 17$ \acounits\ with $0.3$~dex uncertainty. These two results agree within ${\sim} 2\sigma$ of their uncertainties.

Motivated by the dust-inferred $\alpha_\mathrm{CO}$ values derived for Local Group galaxies by \citet{Leroy11} and 27 nearby spiral galaxies by \citet{Sandstrom13}, \citet{Bolatto13} proposed the formula $\alpha_\mathrm{CO} = \alpha_\mathrm{CO,\,MW} f_\mathrm{COF} f_\mathrm{SB}$ in which $\alpha_\mathrm{CO}$ may deviate from the Galactic value $\alpha_\mathrm{CO,\, MW}$ for two reasons: the factor $f_\mathrm{COF}$ accounts for the H$_2$ mass in the outer complex CO-faint (COF) layers where CO is dissociated; and $f_\mathrm{SB}$ accounts for changes in the CO surface brightness (SB) due to temperature and velocity dispersion. The former term may scale as $f_\mathrm{COF} \approx 0.67 \exp{(+0.4 / Z^\prime \Sigma_\mathrm{GMC}^{100})}$ as derived by \citet{Wolfire10}. For the cloud complexes in NGC 6822 ($Z^\prime = 0.2$ and $\Sigma_\mathrm{GMC}^{100} = 1.05$ \Msunperpc) this results in $f_\mathrm{COF} \approx 5$ and thus a similar increase in $\alpha_\mathrm{CO}$ (assuming no change in $f_\mathrm{SB}$ given the similarity of the CO-bright clumps in NGC 6822 and the Milky Way). This prediction is $\sim 4{-}5$ times below our measurement of $\alpha_\mathrm{CO}$ for complexes in NGC 6822. This deviation does not come as a surprise as \citet{Bolatto13} empirically calibrated their formula against the $\alpha_\mathrm{CO}$ values derived by \citet{Leroy11} and \citet{Sandstrom13}, and we already noted the difference of our and \citeauthor{Leroy11}'s results above. The underlying reason for this discrepancy remains unclear, but dust maps of higher fidelity seem necessary to resolve these questions.

As \autoref{f6} has shown, we find different $\alpha_\mathrm{CO}$ values for different complexes with a substantial difference between the two CO-bright Fields 2~\&~4, where $\alpha_\mathrm{CO} \approx 85$ \acounits , and the CO-faint Fields 1~\&~3, where $\alpha_\mathrm{CO} \approx 235 - 570$ \acounits . The difference may be linked to the evolutionary state of the complexes. The two CO-faint complexes (Fields 1~\&~3) with high $\alpha_\mathrm{CO}$ also have prominent \hii\ regions with fewer signatures of compact embedded star formation at $24~\micron$. This might also lead to more dissociating photons and weaker shielding because the gas has been dispersed by stellar feedback, in which case we might expect a larger component of H$_2$ without CO.  By contrast, the other two complexes (Fields 2~\&~4) may be in an earlier state, with embedded star formation and gas more concentrated and better shielded. This could lead to the overall brighter CO emission and the lower conversion factors. While this scenario does make sense ``by eye,'' we might also expect this case to produce CO intensity distributions that differentiate the fields in this way. However, this does not appear to be the case for our four complexes and warrants further investigation with larger samples. 

We also constrain the CO-to-H$_2$ conversion factor via dynamical measurements at small scales of individual dense clumps. If we assume that the dynamical state of the clumps in NGC 6822 and WLM is the same as for those in the Milky Way, then the difference in the apparent ratio of virial mass to luminosity reflects differences in $\alpha_\mathrm{CO}$ as small as a factor of $\lesssim 2$ despite metallicity variations of ${\sim} 1/8 - 1~Z_\odot$ when measured for dense clumps on spatial scales of only a few parsec.

In summary, this highlights the strong dependence of $\alpha_\mathrm{CO}$ on both metallicity and spatial scale. The CO-to-H$_2$ ratio can be very low in low metallicity systems on spatial scales (few $10\,\mathrm{pc} - 1\,\mathrm{kpc}$) of whole molecular clouds to the entire galaxy while it can approach Galactic values inside small ($\sim$pc), dense, well-shielded gas clumps. Accounting for the scale dependence of $\alpha_\mathrm{CO}$ seems to bring previous, apparently differing estimates of the metallicity dependence of $\alpha_\mathrm{CO} \propto Z^{-\gamma}$ together \citep[see also][]{Bolatto13, Schruba13}. Using dust modeling to infer H$_2$ masses, \citet{Lee15} derived $\gamma \approx 1{-}2$ on spatial scales of ${\sim} 10$~pc in the LMC and SMC, while \citet{Leroy11} and \citet{Sandstrom13} derived $\gamma \approx 1.5$ on $\sim$kpc scales in local galaxies. From scaling the SFR by a depletion time to get H$_2$ masses, \citet{Genzel12},  \citet{Schruba12}, and \citet{Hunt15} derived $\gamma \approx 2{-}3$ for entire galaxies in the local and distant universe. Furthermore, we find that $\alpha_\mathrm{CO}$ as measured for individual molecular clouds depends on the cloud's evolutionary state driven by particular dynamical, chemical, and feedback timescales (see \autoref{f6}). Here we suggest that these $\alpha_\mathrm{CO}$ measurements can be brought together when considering the interlinked metallicity and spatial scale dependence of $\alpha_\mathrm{CO}$; but sensitive observations of CO and dust continuum at high resolution for a set of low metallicity galaxies are critical for a definite answer.

\subsection{The Atomic-Molecular Complexes}
\label{subsec:stability}

We can use the derived properties (i.e., size, mass, surface and volume density) of the atomic-molecular complexes and the CO-bright clumps to analyze their ``state of equilibrium'' and determine if they share the same properties as clouds in the Milky Way. The basic idea consists of (numerous) CO clumps residing in an atomic-molecular complex that have to be close to pressure and chemical equilibrium so that the entire gas cloud is in a (quasi) static state. In \autoref{subsec:clumps} we have determined that the pc-sized, CO-bright clumps have properties suggesting that these structures are in gravitational, excitation, and pressure equilibrium, and that their properties are similar to Galactic clumps. Here we extend this analysis to consider the equilibrium state of the ${\sim} 100$ times larger atomic-molecular complexes.

The atomic-molecular complexes have a  (dust-inferred) typical diameter of $ D = 110$ pc and gas mass of $M_\mathrm{gas} = 1.3 \times 10^6$ \Msun, the average surface density is $\Sigma_\mathrm{gas} = 105$ \Msunperpc, and about $30\%$ of their mass is in atomic form and $70\%$ in molecular form. These complexes are surrounded by a diffuse atomic medium with average surface density of $\Sigma_\mathrm{ISM} = 20$ \Msunperpc, as determined from the \mbox{21-cm} observations.

In a static picture, the reservoir of H$_2$ and CO gas has to be in an equilibrium between formation and destruction processes. This requires that H$_2$ and CO are sufficiently shielded from dissociating radiation. In the Milky Way and Magellanic Clouds, H$_2$ molecules require an extinction of $A_\mathrm{V} \approx 0.3$ mag and CO molecules require $A_\mathrm{V} \approx 1.5$ mag \citep[we note that these numbers are rough, first order estimates, they reflect the good agreement among theoretical and observations studies on the magnitude of $A_\mathrm{V}$ but may neglect a minor, still debated metallicity dependence, e.g.,][and Lee et al., ApJ subm.]{Krumholz08, Pineda08, Wolfire10, Glover11, Sternberg14, Lee15}. At solar metallicities, this corresponds to gas columns of $\Sigma_\mathrm{gas} \approx 6$ \Msunperpc\ and $\approx 30$ \Msunperpc , respectively. In NGC 6822, which has ${\sim} 1/5$ solar metallicity, the corresponding gas columns are expected to be $\Sigma_\mathrm{gas} \approx 30$ \Msunperpc\ and $\approx 150$ \Msunperpc , under the assumption that $A_\mathrm{V}$ is linearly proportional to metallicity. The former is satisfied by the atomic component of the atomic-molecular complexes with $\Sigma_\mathrm{atom} \approx 0.3 \times 105$ \Msunperpc\ plus additional shielding from the diffuse atomic medium, and the latter by the total gas column of $\Sigma_\mathrm{atom+mol} \approx 105 + 125$ \Msunperpc\ calculated from the atomic and molecular gas of the complex plus the molecular gas from the embedded CO-bright clumps. In regions of strong radiation fields --- our ALMA survey specifically targets prominent \hii\ regions --- the required shielding columns will need to be somewhat higher, what our derived numbers allow for. In summary, the atomic-molecular complexes have properties (i.e., column densities) conforming with chemical equilibrium. The dynamical state cannot be evaluated as we lack information on the gas (turbulent) velocities of the entire complex.

\subsection{Star-forming Gas Content and Depletion Time}
\label{subsec:sfgas}

We have estimated the global \coone\ luminosity and the \hi\ and H$_2$ mass of the star-forming complexes hosting ${\sim} 2/3$ of the star formation activity in NGC 6822. Using these, we are in a position to discuss the relative abundance of gas, molecular gas, and star formation on large scales. NGC 6822 appears typical of dwarf galaxies in two respects: The timescale for the current star formation rate, $\mathit{SFR} \approx 0.015$~M$_\odot$~yr$^{-1}$, to consume all (atomic and molecular) gas of the entire galaxy is long, ${\sim} 9$~Gyr. Such a long timescale is typical of dwarf galaxies and the outer disks of spiral galaxies \citep[e.g.,][]{Lee09a, Bigiel10b, Schruba11, Huang12, Roychowdhury15}. This is usually interpreted as reflecting the inability of dwarf galaxies to effectively convert their atomic material into dense, cold, star-forming clouds.

NGC 6822 also displays a small CO luminosity relative to its apparent rate of star formation. With a CO luminosity of $L_\mathrm{CO} \approx 6{-}10 \times 10^4$~\Kkmperspc , the whole galaxy compares to only the CO luminosity of a single Milky Way high mass star-forming cloud. In this sense it resembles the SMC \citep{Mizuno01}, which also has $L_\mathrm{CO}$ of order ${\sim} 10^5$~\Kkmperspc\ and a star formation rate of ${\sim} 0.05$ \Msunperyr . In its very high ratio of SFR-to-CO, $\sim 50{-}80$ times higher than in (massive) disk galaxies, it conforms to a broader population of dwarf galaxies \citep[see][]{Schruba12}. This low CO luminosity relative to star formation rate is usually interpreted as a selective destruction of CO molecules, as discussed in the introduction.

In detail, however, our mass budget for NGC 6822 holds a surprise. When focusing on our four star-forming complexes, at face value, the ratio of the recent star formation rate to the apparent gas reservoir --- H$_2$, \hi , or both --- appears high. That means, the apparent consumption time of the star-forming gas is quite short. We compare the integrated H$_2$ mass of our complexes (as inferred from dust modeling) to their star formation rate and find that present day star formation will consume the molecular gas content of the complexes in ${\sim} 360$~Myr. This is much shorter than the typical $1-2$ Gyr molecular gas depletion time seen for massive disk galaxies \citep[e.g.,][]{Bigiel11, Saintonge11b, Schruba11, Leroy13b}. Even including the atomic gas associated with the complexes, this consumption time appears short, with star formation being able to consume the all of the complex's atomic and molecular gas (as inferred from dust) in ${\sim} 510$~Myr (or ${\sim} 760$~Myr if we add the diffuse atomic gas component that we had removed by our background subtraction). At first glance, these ``short'' depletion timescales seem to be in good agreement with estimates for other nearby low mass ($M_\star \approx 10^8 - 10^9$ \Msun), low metallicity ($Z \approx 0.1 - 0.5 Z_\odot$) galaxies \citep{Gardan07, Gratier10a, Bolatto11, Bothwell14, Cormier14, Hunt15, Jameson16}, but are in conflict with the very long depletion times (${\sim} 60$ Gyr) found by \citet{Shi14}. All these studies (with the exception of \citeauthor{Shi14}) suggest that the molecular gas depletion time in low mass, low metallicity dwarf galaxies is a factor $\sim 2 {-} 5$ shorter than in massive disk galaxies.

Here we want to bring forward a line of thought why the depletion time in NGC 6822 (and potentially other dwarf galaxies) may be considerably longer than the ${\sim} 0.5$~Gyr repeatedly stated. Our survey of NGC 6822 has selectively targeted regions of current active star formation; a selection bias that frequently applies to observations of (dwarf) galaxies. However, when considering the time evolution of the gas-star cycle in galaxies, these actively star-forming regions may not be the regions that host most star-forming gas. Especially if the formation timescale of star-forming gas clouds in much longer than the star formation process itself, then a significant amount of the star-forming gas is in ``quiescent'' clouds that may have been missed by surveys that selectively targeted regions of active star formation. This caveat especially applies to studies of nearby galaxies with large angular extent on the sky.

The biasing in the gas depletion time for small apertures selectively targeting star-forming or gas-rich regions has been studied observationally by \citet{Schruba10} and theoretically by \citet{Kruijssen14a}. Here we utilize the \citeauthor{Kruijssen14a} model to estimate the bias in the depletion time when focusing selectively on star-forming regions as compared to a complete survey of the entire galaxy. In detail, we use their Eq.~16 which is
\begin{eqnarray}
\frac{t_\mathrm{dep,\,star}}{t_\mathrm{dep,\,gal}} =
\frac{\beta \frac{t_\mathrm{over}}{t_\mathrm{gas}} 
{[1 + (\beta -1 ) \frac{t_\mathrm{over}}{t_\mathrm{gas}}]}^{-1} +
\frac{t_\mathrm{star}}{t_\mathrm{tot}}
{(\frac{l_\mathrm{ap}}{\lambda})}^2}
{1 + \frac{t_\mathrm{star}}{t_\mathrm{tot}}
{(\frac{l_\mathrm{ap}}{\lambda})}^2}
\label{eq:bias}
\end{eqnarray}
and adopt the following parameters: lifetime of the complexes $t_\mathrm{gas} = 35$ Myr \citep[one dynamical timescale at the location of our survey fields;][]{Weldrake03}; duration of star formation $t_\mathrm{over} = 5$ Myr \citep[age spread of massive stars from HST;][]{ODell99}; visibility timescale of (H$\alpha$) star formation tracer $t_\mathrm{star} \equiv t_\mathrm{H \alpha} + t_\mathrm{over} = 10$ Myr \citep[with $t_\mathrm{H \alpha} = 5$ Myr from stellar synthesis modelling;][and D.~T.\ Haydon et~al., in preparation]{Leroy12}; total time range $t_\mathrm{tot} \equiv t_\mathrm{gas} + t_\mathrm{star} - t_\mathrm{over}$; separation of SF regions $\lambda = 400$ pc (estimated from H$\alpha$ map); fraction of gas decrease during SF process $\beta = 1$; and the aperture size $l_\mathrm{ap}$ as independent variable. We note that these parameters are rough estimates; a detailed derivation and analysis of uncertainties is beyond the scope of this paper and will be presented in forthcoming papers.

\begin{figure}[t]
\epsscale{1.09}
\plotone{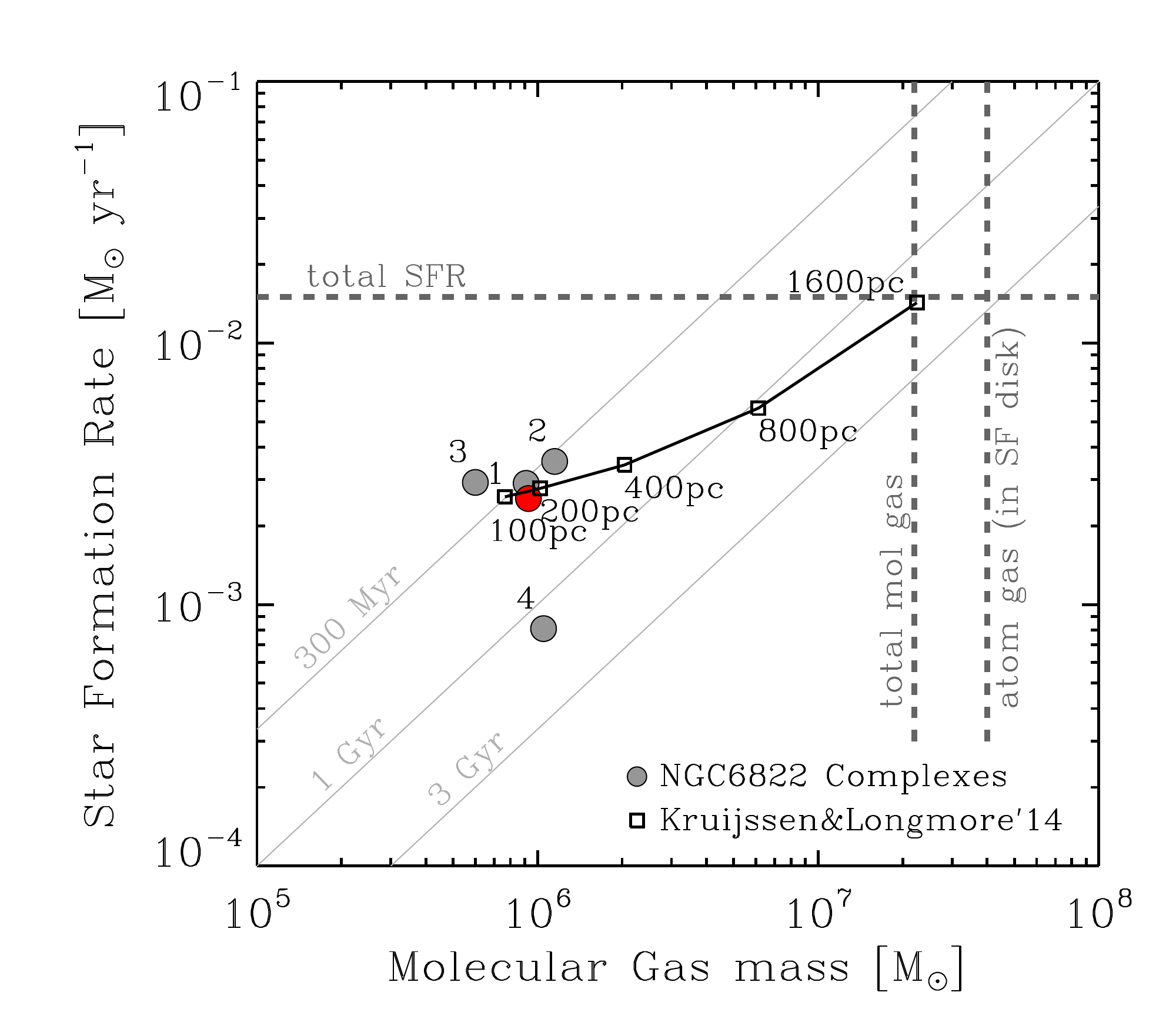}
\caption{Molecular gas depletion time, $\tau_\mathrm{depl}$, the ratio of molecular gas mass and star formation rate, as function of spatial scale: ranging from individual star-forming complexes ($d \approx 100$ pc) to the entire star-forming disk of NGC 6822 ($d \approx 1600$ pc). The \citet{Kruijssen14a} model provides an estimate of the bias in $t_\mathrm{dep}$ when selectively focusing on star-forming regions (black curve). We infer that globally, NGC 6822 may have $t_\mathrm{dep} \approx 2$ Gyr (similar to massive disk galaxies) and molecular gas mass $M_\mathrm{mol} \approx 2 \times 10^7$~\Msun .
\label{f10}}
\end{figure}

\autoref{f10} quantifies the bias in the molecular gas depletion time as derived by the \citet{Kruijssen14a} model. It shows the molecular gas mass (as inferred from the dust) and SFR measurements for the four complexes (gray circles) and their mean average (red circle), as well as the model-predicted scale-dependence of the bias (black curve) for the adopted parameters. In detail, the model predicts how $M_\mathrm{mol}$ and $\mathit{SFR}$ scale with aperture size, $l_\mathrm{ap}$,
\begin{eqnarray}
M_\mathrm{mol,\,model} &=& \Sigma_\mathrm{mol,\,gal} \times
\left( l_\mathrm{ap}^2 + \frac{t_\mathrm{over}}{t_\mathrm{star}} \lambda^2 \right)
\label{eq:mmol}\\
\mathit{SFR}_\mathrm{model} &=& 
\frac{M_\mathrm{mol,\,model}}{t_\mathrm{dep,\,gal}} \times
{\left( \frac{t_\mathrm{dep,\,star}}{t_\mathrm{dep,\,gal}} \right)}^{-1}
\label{eq:sfr}
\end{eqnarray}
\noindent In \autoref{eq:mmol}, the first term in parentheses reflects the `random' background population of clouds and star-forming regions, whereas the second term reflects the contribution of the region that the aperture is focussed on, with the ratio $t_{\rm over}/t_{\rm star}$ indicating the probability that the region contains a representative gas reservoir. In \autoref{eq:sfr}, the bias of the gas depletion time from \autoref{eq:bias} is used to convert the molecular gas mass to an SFR. The model is constrained at small aperture size ($l_\mathrm{ap} \approx 100$ pc) by the average $M_\mathrm{mol}$ and $\mathit{SFR}$ for the star-forming complexes and at large aperture size ($l_\mathrm{ap} \approx 1.6$ kpc) by the total $\mathit{SFR}$ over the star-forming disk of NGC 6822 (a rectangle of $1.1 \times 2.4$ kpc$^2$). These boundary conditions fully constrain the galaxy's molecular gas depletion time and total molecular gas mass which estimate as $t_\mathrm{dep,\,gal} \approx 2$ Gyr and $M_\mathrm{mol\\,\,gal} \approx 2 \times 10^7$ \Msun , respectively. This implies a molecular gas fraction of ${\sim} 0.35$ over the star-forming disk or ${\sim} 0.15$ for all of NGC 6822; values that are in good agreement with estimates for other dwarf galaxies \citep{Bothwell14, Hunt15}.

If all molecular gas is in massive complexes of ${\sim} 10^6$ \Msun\ (as estimated for our four targeted complexes), then we expect ${\sim} 20$ such complexes in NGC 6822. It remains unclear though if this expectation holds. \citet{Gratier10a} identified 15 CO-bright clouds in their IRAM 30-m CO map (covering about half the star-forming disk) but those have typically a lower mass of ${\sim} 10^5$ \Msun . Alternatively, (the dust component of) the predicted clouds should be visible in the \emph{Spitzer} and \emph{Herschel} infrared maps, but those do not provide a conclusive answer due to confusion with Galactic foreground emission. On the other hand, our estimate of a long depletion time is in agreement with theoretical predictions that on large ($\gtrsim 1$~kpc) scales, star formation and the amount of molecular gas may show a nearly constant ratio for metallicity $Z \gtrsim 0.1$ Z$_\odot$ because star formation and H$_2$ formation have similar density and shielding requirements \citep[e.g.,][]{Krumholz11, Glover12a, Glover12b}. 

The presence or absence of a large CO-faint H$_2$ component pivots on joint analysis of the dust and gas data. This process has a number of subtleties and uncertainties, which are discussed in detail in, e.g., \citet{Leroy07, Leroy11} and \citet{Sandstrom13}. This analysis is not the focus of this paper; given the apparent discrepancy between our results and those of other recent analyses, we defer a detailed commentary until we are able to achieve a better determination of the dust content and analyze the whole set of Local Group galaxies in a self consistent way. Furthermore, a large-scale survey of neutral or ionized carbon line emission could verify the existence of a large CO-faint H$_2$ component.

Another explanation leading to a short depletion time could be invoked in time rather than space: if NGC 6822 recently experienced a large burst of star formation then we might observe this burst in star formation tracers but the reservoir for the burst might have been dispersed by feedback. This scenario also does not appear viable. The star formation rate in NGC 6822 has been roughly constant over the past $\sim 400$ Myr \citep[with variations of a factor $\lesssim 2$ over durations of a few 10's of Myr;][]{Efremova11} and the galaxy displays a typical specific star formation rate ($\mathit{SFR} / M_\star$). There is no evidence that NGC 6822 
\mbox{has recently undergone a galaxy-wide starburst.}

\section{Summary}
\label{sec:summary}

We present the first ALMA maps of the molecular gas in the Local Group star-forming dwarf irregular galaxy NGC 6822. We observe five fields, detecting \cotwo\ in the four fields showing active high mass star formation. These regions together encompass ${\sim} 2/3$ of the star formation activity in the galaxy and within them we resolve ${\sim} 150$ compact CO clumps. We measure the properties of these clumps at a resolution of 2~pc, a sharpness of view previously only achievable in the Milky Way. We find the bright CO emission to correlate well with the location of PAH emission seen in {\em Spitzer} $8~\micron$ imaging. The relationship of CO to $24~\micron$ emission is less straightforward at these high resolutions, and the correspondence with H$\alpha$ on these scales is poor. Compared to high mass Galactic star-forming clouds, the regions that we observe in NGC 6822 have a narrower distribution of CO intensities on average and are ${\sim} 30\%$ less bright.

In agreement with \citet{Rubio15} but considering a $15$ times larger sample of clumps, we show that CO emission from low metallicity galaxies originates from very compact, bright regions with small radii of $\sim 2{-}3$~pc and narrow line widths of ${\sim} 1$ \kmpers . The macroscopic properties of these CO bright clumps are very similar (with possible offsets $\lesssim 2$) to CO-bright structures within the Milky Way's Perseus arm and WLM. The only noticeable difference between NGC 6822 and the Galaxy on these scales appear to be ${\sim} 30\%$ lower surface brightnesses and a factor of ${\lesssim} 2$ higher virial masses or virial ratios for fixed (CO-inferred) luminous masses. These differences can be resolved by either sub-thermal excitation of the observed \mbox{CO(2-1)} transition, a CO-to-H$_2$ conversion factor ${\sim} 2$ times larger than the standard galactic value, or due to the additional weight (i.e., surface pressure) of the increasing shielding layers. Within their confidence level all three explanations are viable. One of our main results is that the (well-shielded) CO-bright clumps in NGC 6822 have similar physical properties such as size, column and volume density, and dynamical equilibrium state as equally-sized structures in the Milky Way and largely unaffected by the low (${\sim} 1/5$ solar) metallicity of NGC 6822.

By modeling the infrared emission of dust, we infer total gas masses of the atomic-molecular complexes hosting the CO-bright clumps of $\sim 10^6$ \Msun . About $30\%$ of their mass is in atomic form and $70\%$ in molecular form. The inferred surface densities are sufficiently high to shield H$_2$ within the entire complexes and CO within the dense clumps. This disparate distribution of H$_2$ and CO leads to a strong dependence of the CO-to-H$_2$ conversion factor, $\alpha_\mathrm{CO}$, on both metallicity and spatial scale. The CO-to-H$_2$ ratio can be very low in low metallicity systems on spatial scales (few $10\,\mathrm{pc} - 1\,\mathrm{kpc}$) of whole molecular clouds to the entire galaxy while it can approach Galactic values inside small ($\sim$pc), dense, well-shielded gas clumps.

At the current star formation rate, the molecular gas component of the complexes will be exhausted within ${\sim} 360$ Myr; a timescale in agreement with previous measurements of other nearby, low mass galaxies. However, we caution that this short timescale may not apply to the entire galaxy since our survey targeted selectively regions of active star formation. We apply the theoretical model by \citet{Kruijssen14a} to estimate the bias in the depletion time for small apertures focused on star-forming regions and find that the global molecular gas depletion time may be as long as ${\sim} 2$ Gyr and thus similar to massive disk galaxies. This analysis also suggest a significant population of quiescent molecular clouds and a considerable molecular gas fraction of ${\sim} 0.35$ within the star-forming disk of NGC 6822 or ${\sim} 0.15$ the entire galaxy. The alternative explanation to explain the short depletion times -- a recent burst in star formation -- does not seem to apply to NGC 6822. We caution that these results require detailed follow-up but they raise the possibility that low metallicity dwarf galaxies may harbor significantly more molecular gas than inferred by previous surveys and that low and high mass galaxies may share similar molecular gas depletion times as has been proposed in several theoretical models \citep[e.g.,][]{Krumholz11, Glover12a, Glover12b}. A future homogeneous analysis of gas and dust tracers of all Local Group galaxies will provide a definitive answer.

Beyond the mystery of the complex-scale accounting, this work raises several prospects for fruitful follow-up. The good correspondence between high resolution $8~\micron$ and CO emission reinforces the potential, already highlighted by \citet{Sandstrom10} and \citet{Gratier10a}, for PAH signatures including those observed by WISE at $12~\micron$ to trace molecular gas in the low metallicity galaxies. The compact, confined nature of the CO emission in NGC 6822 agrees with the physical picture that a lack of dust shielding pushes CO emission deep into molecular clouds at low metallicity; a direct, pc-scale comparison of the distributions of dust and CO could measure whether the threshold for CO emission is indeed the same in dwarf galaxies and the Milky Way \citep[see, e.g.,][]{Lee15}. Finally, the highly structured nature of the emission in our fields offers the possibility to employ a combination of CO and \hi\ kinematics to probe the kinematics of gas in atomic-molecular complexes across the whole range of scales present in one of our fields, in the process 
\mbox{offering additional constraints on the dynamical mass.}

\acknowledgments

The authors thank the anonymous referee for a careful and constructive report that improved this paper. We are very grateful to Pierre Gratier for sharing the IRAM \mbox{30-m} data, Ioannis Bagetakos for his reduction of the Little Things VLA data, and Maud Galametz and Sue Madden for the Herschel data and dust modeling results. We would further like to thank Anand Crossley and Sarah Wood from NAASC for help regarding the execution of the ALMA observations and initial data reduction. AS would like to thank Joe Mottram and Frank Israel for insightful discussions. ADB acknowledges partial support from NSF-\mbox{AST0955836} and NSF-\mbox{AST1412419}. JMDK gratefully acknowledges financial support in the form of a Gliese Fellowship and an Emmy Noether Research Group from the Deutsche Forschungsgemeinschaft (DFG), grant number \mbox{KR\,4801/1-1}. FB acknowledges support from DFG grant \mbox{BI\,1546/1-1}. The work of WJGdB was supported by the European Commission (grant FP7-PEOPLE-2012-CIG \#333939). Astrochemistry in Leiden is supported by the European Union \mbox{A-ERC} grant 291141 CHEMPLAN, by the Netherlands Research School for Astronomy (NOVA), and by a Royal Netherlands Academy of Arts and Sciences (KNAW) professor prize. 

This paper makes use of the following ALMA data: ADS/JAO.ALMA\#2013.1.00351.S (PI. A.~Schruba). ALMA is a partnership of ESO (representing its member states), NSF (USA) and NINS (Japan), together with NRC (Canada) and NSC and ASIAA (Taiwan), in cooperation with the Republic of Chile. The Joint ALMA Observatory is operated by ESO, AUI/NRAO and NAOJ. The National Radio Astronomy Observatory is a facility of the National Science Foundation operated under cooperative agreement by Associated Universities, Inc.


\end{document}